\documentclass{aa}  
\usepackage{graphicx}
\usepackage{txfonts}
\begin{document}
   \title{The extragalactic optical-infrared background radiation, \\ 
  its time evolution and the cosmic \\  photon-photon opacity
}

\titlerunning{Background radiations and the cosmic photon-photon opacity}


   \author{A.~Franceschini\inst{1}, G.~Rodighiero\inst{1}, M.~Vaccari\inst{1}
          }

   \offprints{A. Franceschini}

   \institute{Dipartimento di Astronomia, Universita' di Padova,
I-35122 Padova, Italy\\
              \email{alberto.franceschini@unipd.it}
             }

   \date{Received February 29, 2008; accepted May 12, 2008}

 
  \abstract{
   The background radiations in the optical and the infrared
  constitute a relevant cause of energy loss in the propagation of
  high energy particles through space. In particular, TeV observations
  with Cherenkov telescopes of extragalactic sources are influenced by
  the opacity effects due to the interaction of the very high-energy
  source photons with the background light.  
   With the aim of assessing with the best possible detail these
  opacity terms, we have modelled the extragalactic optical and
  infrared backgounds using available information on cosmic sources in
  the universe from far-UV to sub-millimeter wavelengths over a wide
  range of cosmic epochs.  
 We have exploited the relevant cosmological survey data -- including
  number counts, redshift distributions, luminosity functions -- from
  ground-based observatories in the optical, near-IR, and
  sub-millimeter, as well as multi-wavelength information coming from
  space telescopes,  HST, ISO and Spitzer. Additional constraints have
  been used from direct measurements or upper limits on the
  extragalactic backgrounds by dedicated missions (COBE). All data
  were fitted and interpolated with a multi-wavelength backward
  evolutionary model, allowing us to estimate the background photon
  density and its redshift evolution.  
   From the redshift-dependent background spectrum, the photon-photon
  opacities for sources of high-energy emission at any redshifts were
  then computed.  The same results can also be used to compute the
  optical depths for any kind of processes in the intergalactic space
  involving interactions with background photons (like scattering of
  cosmic-ray particles). 
   We have applied our photon-photon opacity estimates to the analysis
  of spectral data at TeV energies on a few BLAZARs of particular
  interest.   The opacity-corrected TeV spectra  are entirely
  consistent with standard photon-generation processes and show photon
  indices in any case steeper than $\Gamma_{intrinsic}=1.6$.    
Contrary to some previous claims, but quite in agreement with Aharonian et al. (2006), we find no evidence for any truly diffuse background components in addition to those from resolved sources.
We have tested in particular the effects of a photon background
  originating at very high redshifts, like would be the emissions by a
  primeval population of Population III stars around $z\sim 10$.  We
  could not identify any opacity features in our studied BLAZAR
  spectra consistent with such an emission and put a stringent limit
  on such diffuse photon intensity of $\sim 6$ nW/m$^2$/sr between 1
  and 4 $\mu$m.   
 
   TeV observations of BLAZARs are consistent with background
   radiations contributed by resolved galaxies in the optical and IR,
   and exclude prominent additional components from very high-z
   unresolved sources.
}

   \maketitle


%

\section{Introduction}

The extragalactic background radiation at various electro-magnetic frequencies constitutes a fundamental source of opacity for the propagation of high-energy cosmic-ray (CR) particles and photons throughout space-time. The origin of this opacity depends on the physical processes involved. Protons for example, which are the most numerous population of cosmic-rays in the local universe, interact with background photons by generating neutrons and charged mesons ($p^+$ and $\mu^+$), which initiate a decay chain eventually producing electron and muon neutrinos and anti-neutrinos (Berezinsky and Zatsepin 1969; Stecker 1973; Stanev 2004). These interactions of CR and background photons are currently considered to be the main source of the local high-energy neutrino background (\textit{cosmogenic neutrinos}) that may be detected by forthcoming experiments (e.g. AMANDA, see Achterberg et al. 2007, and the IceCube neutrino observatory, see Achterberg et al. 2006).

As first pointed out by Nikishov (1962) and further discussed in a variety of papers (Gould and Schreder 1966; Stecker 1969; Stecker et al. 1992; among others), very high-energy photons also suffer opacity effects by photon-photon interactions with local backgrounds and pair production. As a result of the large number of photons (400 $cm^{-3}$) in the cosmic microwave background (CMB), any photons with energy $\epsilon>100$ TeV traveling in space have a very short mean free path before decaying, such that extragalactic sources are essentially undetectable above this energy. Below this threshold, source detectability depends essentially on its cosmic distance, due to the photon opacity induced by extragalactic backgrounds other than the CMB. 

All this is currently the subject of extensive investigation with imaging atmospheric Cherenkov telescopes, now operating between a few tens of GeV to tens of TeV (e.g. HEGRA, HESS, MAGIC): particularly debated is how much the TeV observations of distant BLAZARs are, or are not, consistent with the photon-photon absorption effect (e.g. Aharonian et al. 2006; Albert et al. 2006).
Indeed, as originally proposed by Stecker et al. (1992), the observations of absorption features in the high-energy spectra of distant BLAZARs may be used to place relevant constraints on the spectral intensity of background radiation (see also Stecker and De Jager 1993, Dwek and Slavin 1994, Stanev and Franceschini 1998, Renault et al. 2001, Mazin and Raue 2007).

Thus a whole new field of astro-particle physics now benefits from the observational capabilities of very high energy photons by air Cherenkov telescopes, the various X-ray observatories in space, the forthcoming gamma-ray space observatory \textit{GLAST}, and the currently improved knowledge of the photonic cosmic background radiation. Similarly interesting prospects now exist for the detection of high-energy neutrinos by various ongoing neutrino experiments.

So far these analyses have mostly targeted nearby objects, particularly the bright BLAZARs MKN 421 and MKN 501, and have considered the absorption effects produced by the local photonic backgrounds.
However, the present-day possibilities of detecting high-energy neutrinos and photons from sources at cosmological distances ($z>>0.1$) require that a detail account is made of how the photon backgrounds evolve in cosmic time along the source's lines-of-sight.

The major contributor to the background photon flux, the CMB, has been extremely well characterized both spectrally (Mather et al. 1994) and spatially (Hinshaw et al. 2007), and has been shown to be a very isotropic pure black-body spectrum. Measurements of the CMB from space (COBE and W-MAP) have been eased by the far dominance of this component over the foreground emissions by the Galaxy and interplanetary dust (IPD). In consideration of its origin in the primeval plasma at $z\sim1000$, the evolution of this radiation in cosmic time is also determined, the photon number density $n_\gamma$ simply scaling with $z$ as $n_\gamma\propto(1+z)^3$.

On the contrary, the situation for the other extragalactic radiations of interest, the cosmic far-infrared background (CIRB) and the extragalactic background light in the UV-optical-near-IR (EBL), is quite different. On one side, their observation is hampered or even prevented by the presence of intense foreground emission over most of the wavelength range (Hauser et al. 1998). In addition, these radiations are progressively generated by galaxies and active nuclei (AGN) during most of the Hubble time, particularly below $z=1$, so that the evolution of their photon number density is a very complex function of time and frequency. 
Some attempts have been made to represent this evolution with simple parametric expressions. In particular, Stecker, Malkan and Scully (2006) have studied the time evolution of the background radiation using simple power-law approximations to the evolution rates as a function of redshift.
However, the real situation revealed by recent observations (particularly by the Spitzer Space Telescope) is that of a more complicated evolution of galaxy emissions driven by a variety of physical processes and cosmic sources, including star-formation, galaxy merging, nuclear AGN emissions and evolving stellar populations, dust extinction and re-radiation. There is also evidence that the evolution rates of the cosmic source emissivities in the optical/near-IR should be different from those in the infrared (e.g. Franceschini et al. 2001).

Fortunately, an enormous amount of new data at all UV-optical and IR-millimetric wavelengths have been recently obtained with astronomical observatories on ground and in space to characterize such an evolution. 
By exploiting all these new data, the present paper is devoted to a complete and detailed modelling of the CIRB and EBL local backgrounds and of their evolution with cosmic time, to provide the most accurate assessment of the number density of background photons as a function of photon energy and redshift. At variance with some previous analyses, we develop separate models of the optical/near-IR and far-IR/sub-mm evolution of galaxies and AGNs, to account for the very different astrophysical processes dominating sources in the two spectral regimes.
We do not consider here other background components, like the X-ray or radio backgrounds, not involved in the cosmic opacity effects.

Evolutionary effects in the infrared background photon density do not affect the $\gamma-\gamma$ opacity in distant BLAZARs, whose spectra are rather influenced by the evolution of the optical-UV background light (Sect. \ref{conc}). The evolution of the CIRB background, instead, is critical for various other reasons, like the production of high-energy neutrinos by interactions of cosmic-ray protons with IR background photons (Stanev et al. 2006), or the photo-excitation of molecular or atomic species in high-redshift media.

All this information will be used to calculate here the cosmic opacities for photon-photon interactions and in a future paper for the CR proton-photon decay.
Our approach consists of establishing the minimum level of background photons based on known sources (galaxies and AGNs) and to assume that these are the only background sources. A comparison with cosmic opacity data, e.g. from BLAZAR TeV observations, will show if such an assumption is correct or whether additional components of truly diffuse photons, for example produced at very high redshifts, do exist.

The paper is organized as follows. Section 2 summarizes direct observations and constraints on the diffuse background light in the optical and infrared. These data establish a benchmark against which evolutionary models of cosmic sources should compare.
In Section 3 we briefly report on multi-wavelength models of galaxy and AGN evolution fitting the whole variety of existing data, including the local background spectral intensity as well as data on distant and high redhift sources. For reasons mentioned above, the models are split into two parts, one considering the far-IR and sub-mm spectral domain, the other the UV/optical/near-IR part. 
Section 4 presents our results for the cosmic photon-photon opacity (the calculation of the proton mean free path for collision with background photons will be performed in a future paper).
Section 5 shows a few applications to recent TeV observations of distant BLAZARs and discusses some important constraints emerging on the background radiation.      Section 6 summarizes our results.

All quantities are computed in this paper assuming a geometry for the Universe with $H_0=70$ km s$^{-1}$ Mpc$^{-1}$, $\Omega_m=0.3$, $\Omega_\Lambda=0.7$.
We indicate with the symbol $S_{24}$ the flux density in Jy at 24 $\mu$m or the luminosity in erg/s/Hz (and similarly for other wavelengths).


\section{The local background spectral intensity}

The cosmological background radiation, a fundamental cosmological observable, is the integrated mean surface brightness of the sky due to resolved and unresolved extragalactic sources in a given waveband. 
For our present purposes, the infrared and optical backgrounds (CIRB and EBL) provide important information for two reasons. On one side, they establish the number density $n_\gamma$ at $z=0$ of photons with frequency $\nu$ through the relation $n_\gamma=4\pi I_0(\nu)/(ch\nu)$, where $I_0(\nu)$ is the background intensity and $\epsilon=h\nu$ the photon energy. In addition, the knowledge of the total background $I_0(\nu)$ sets an essential constraint on the integrated emission of faint distant galaxies, particularly in the case when, for technological limitations, observations at faint flux levels in a given waveband are not possible (as it is still partly the case in the IR/sub-mm domain).

\subsection{Constraints on the UV-optical background}
\label{optbck}

Extragalactic background light at UV and optical wavelengths (say 2500 \AA \ to 1 $\mu$m) is expected to be produced by stellar nucleosynthesis at redshifts z$<$7, and by optically bright (mostly type-I) AGNs.
Unfortunately, the presence of intense foreground emission, orders of magnitude larger than the expected extragalactic signal, makes any direct measurements extremely difficult.
A variety of different approaches have been attempted to measure the optical background, including a pioneering one by Mattila (1976) trying to isolate it by the difference of the signals in and out the lines-of-sight to high Galactic latitude "dark" clouds acting as a "blank screen," spatially isolating all foreground contributions to the background radiation. Toller (1983) later attempted to avoid both atmospheric and Zodiacal foreground by using data taken by the Pioneer 10 spacecraft at 3 AU from the Sun, beyond the Zodiacal dust cloud. 
All such attempts provided us with only upper limits to the optical background. Among the fundamental difficulties was the estimation and subtraction of the Galactic stellar light reflected by high latitude dust (the IRAS "cirrus"). 

To overcome these difficulties, an alternative approach has been followed, among others, by Madau \& Pozzetti (2000). They exploited the convergence of deep galaxy counts in the HDF 
in all optical UBVIJHK bandpasses to estimate a lower limit to the background by integrating the counts over the observed wide magnitude intervals.
These data are reported in Fig. \ref{bkg}, and correspond to an integrated optical background of around 10 nW/m$^2$/sr.

More recently, Bernstein et al. (2002) have re-attempted to estimate the optical background by subtracting from the total signal measured by HST the contributions of the Zodiacal light and the diffuse galactic light. They obtain background levels that are typically factors $\sim$3 larger than those estimated from the direct HST counts, which would mean that either a truly diffuse background (of unknown origin) exists, or the HST counts missed 80\% of the light from faint galaxies, which would be surprising considering the depth of the HST HDF images on which the Madau estimate is based.
Some difficulties and inconsistencies in the Bernstein et al. (2002) measurement are discussed by Mattila (2003), particularly concerning the modelling of the diffuse Galactic light and the evaluation of the terrestrial airglow for ground-based observations. A later revision by Bernstein (2007) has demonstrated that the original analysis strongly underestimated the systematic errors and that the background estimates have to be considered as upper limits. 

For all these reasons, we will refer in the following to the determination based on the deep HST galaxy count estimates as the optical/near-IR background.

\subsection{The near-infrared background}
\label{irbck}

The most extensive effort to measure the local extragalactic background radiation in the infrared has been performed with the dedicated COBE mission (see Hauser and Dwek 2001 for a review). 
COBE observed the background light with two instruments between 1 and 1000 $\mu$m. These include the presence of bright foreground (Zodiacal scattered light and interplanetary dust (IPD) emission, Galactic starlight, high Galactic latitude "cirrus" emission), but also two relatively cleaner spectral windows with minima in the foreground emission:
the near-IR (2-4 $\mu$m) and the sub-mm (100-500 $\mu$m) cosmological windows, where redshifted photons by the two most prominent galaxy emission features, the stellar photospheric peak at $\lambda \sim 1\ \mu$m and the dust re-radiation peak at $\lambda \sim 100\ \mu$m, are detectable.

In the near-IR (1.2 and 3.5 $\mu$m), detections of background light signals have been reported by various groups at a level of several tens of nW/m$^2$/sr, after subtraction of the stellar contribution (Gorjian et al. 2000, Wright \& Reese 2000, Cambresy et al. 2001, Dwek et al. 2005), but with substantial  uncertainty due to the poorly understood contribution of the Zodiacal light, and very limited possibility to check their isotropy (Hauser \& Dwek 2001). 
The existence of a bright extragalactic background in the near-IR was also inferred from the analysis of data by the IRTS mission (Matsumoto et al. 2005), with values at 2.2 and 3.5 $\mu$m higher than the COBE inferred values. At shorter wavelengths, the 
Matsumoto et al. and Xu et al. (2002) results continue to rise steeply to $\sim 65$ nW/m$^2$/sr, even above the 95\% confidence-level COBE upper limits, with a spectrum very reminiscent of that of the Sun-scattered Zodiacal light. Once the the galaxy background is subtracted from the total IRTS background, the remaining excess flux from 1 to 4 $\mu$m amounts to approximately (Kashlinsky 2006)
\begin{equation}
I_{excess} \simeq (29\pm 13)\ nW/m^2/sr .
\label{FPIII}
\end{equation}

By comparison with the upper limits to the background light in the optical, this large near-IR excess seems to imply a strong discontinuity at $\sim1\ \mu$m, which has given rise to interpretations in terms of a redshifted UV emission by $z\simeq 10$ primeval (Population III) stars, e.g. by Salvaterra and Ferrara (2003), Kashlinsky et al. (2004), Kashlinsky (2006).  However, a recent analysis of this near-IR excess by Mattila (2006) has found that, after appropriate Zodiacal-light subtraction, there is no evidence for a step in the diffuse light at $1\ \mu$m. Furthermore, Mattila has rised the possibility that the near-IR excess is due to insufficient Zodiacal-light subtraction (see also Dwek et al. 2005 and Thompson et al. 2007).
In the following we will adopt for the near-IR background the values derived from direct integration of the ultradeep HST and Spitzer counts, as a reference. The occurrence of a truly diffuse additional component will be discussed in the light of the constraints imposed by the observations of TeV BLAZAR spectra (Sect. \ref{PIII}).

\begin{figure*}[!ht]
\centering
\includegraphics[angle=0,width=0.9\textwidth,height=0.7\textwidth]{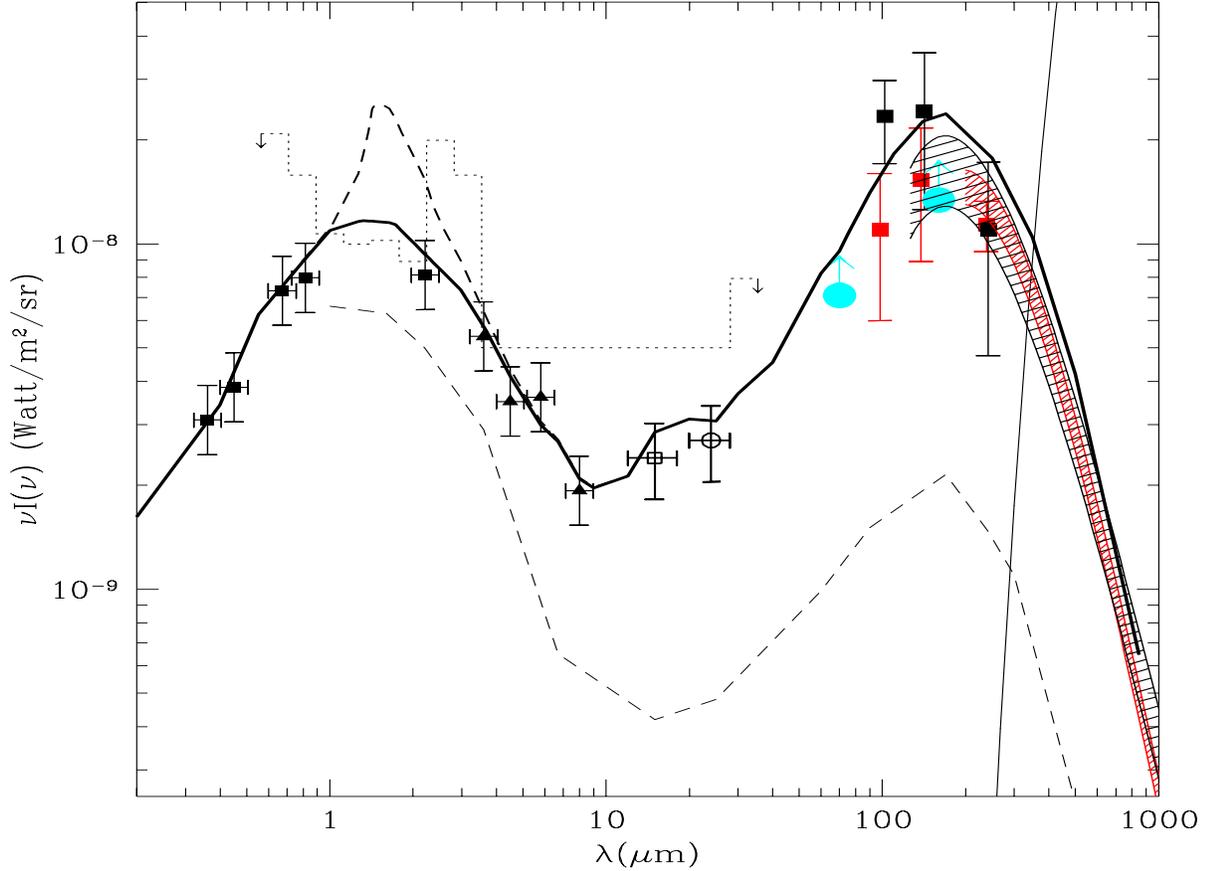}  
\caption{The Cosmic Infrared Background (CIRB) spectrum as measured by independent
groups in the all-sky COBE maps (e.g. Hauser et al. 1998), and the estimates of the optical extragalactic background based on ultradeep integrations by the HST in the HDF (Madau \& Pozzetti 2000). 
The three lower datapoints in the far-IR are from a re-analysis of the DIRBE data by Lagache et al. (1999), the shaded areas from Fixsen et al. (1998) and Lagache et al. The two lower limits at 70 and 160 $\mu$m (green filled circles) come from a stacking analysis of Spitzer data bu Dole et al. (2006).
The two mid-IR datapoints at 15 and 24 $\mu$m are the resolved fraction of the CIRB by the deep ISO surveys IGTES (Elbaz et al. 2002) and Papovich et al. (2004). 
The four triangle datapoints are from the integration of the IRAC galaxy counts by Fazio et al. (2004), except the one at 8 $\mu$m which comes from our own analysis (see text).
The dotted histograms are limits set by TeV cosmic opacity measurements (Stanev and Franceschini 1998).
The thick line is the predicted CIRB spectrum from the multi-wavelength reference model for galaxy evolution discussed in the present paper. The lower dashed line is the expected intensity based on the assumption that the local IR emissivity of galaxies observed by IRAS does not evolve with cosmic time: the distance of the two curves, much larger in the far-IR than it is in the optical/near-IR, illustrates the different evolutionary effects in the two waveband intervals and due to the completely different photon-generating processes. The dashed line corresponds to the inclusion of an excess background intensity from unresolved sources (e.g.  Population III stars), that we have considered, and still marginally consistent with our analysis of TeV BLAZAR spectra, see Sect. \ref{PIII}. The maximal total intensity of this additional background is $I_{bol}\simeq 6$  nW/m$^2$/sr.
}
\label{bkg}
\end{figure*}

The datapoints at $\lambda=3.6$ to 5.8 $\mu$m reported in Fig. \ref{bkg} (triangles) correspond to the integrated signal of all galaxies detected by Spitzer/IRAC as reported by Fazio et al. (2004), derived considering the strong convergence in the counts observed at the faintest fluxes reached by these observations.
On the other hand, because of the uncertain shape of the counts by Fazio et al. at the bright fluxes (where some effects of the local large-scale structure are apparent), the value at 8 $\mu$m has been inferred by us from our own re-assessment of the 8 $\mu$m number counts (see Sect. \ref{nirmod}).

\subsection{The far-infrared background (CIRB)}

The 100-500 $\mu$m long-wavelength spectral region is the only one where an extragalactic signal has been reliably identified. An isotropic background has been estimated by Puget et al. (1996), Fixsen et al. (1998) and Lagache et al. (2000) in COBE/FIRAS spectral maps. 
This detection has been confirmed by various other groups in two broad-band COBE channels at $\lambda = 140$ and $240 \mu$m (Hauser et al. 1998; Schlegel et al. 1998), based on accurate calibration of the COBE/DIRBE maps and detailed spatial models of the foreground values exploiting their spatial variations.

Finkbeiner, Davies \& Schlegel (2000), after a very delicate subtraction of the far dominant Galactic and IPD foregrounds,  found an isotropic signal at 60 and 100 $\mu$m with intensities at the level of $\sim 30\ 10^{-9}\ W\ m^{-2} \ sr^{-1}$. This controversial result was not confirmed by later analyses (see Puget \& Lagache 2001 for a critical assessment).

\section{Photon emissions by cosmic source populations in the optical and infrared}
\label{model}

For our modeling of the photon generation by cosmic sources we have adopted an empirical approach based on the extrapolation back in cosmic time of the locally observed galaxy's luminosity functions. We have preferred this to the alternative scheme of developing a more physical model based on \textit{first principles}, e.g. on the current hierarchical clustering paradigm (see Primack et al. 2005), because the backward empirical scheme allows a link that is more direct, more flexible and closer to the observational data. In addition, it allows a more natural continuity of the past population emissivity with the local background radiation intensity.

This empirical modeling over such a wide wavelength range from far-UV to the millimeter requires some \textit{ad hoc} prescriptions which are different from one waveband to another, due to the widely different physical processes responsible for the photon production. Specifically, galaxies radiate from the UV to the near-IR through the photospheric emission of their constituents stellar populations, while at longer wavelengths photons are produced by the galaxy's diffuse dusty medium.

\subsection{Near-infrared source populations and their cosmological evolution}
\label{nirmod}

To describe the evolutionary properties of source populations in the near-IR (1-10 $\mu$m) we have adopted the model reported by Franceschini et al. (2006, henceforth AF06).   We defer to that paper for details on this modelling and report here only the main facts.

AF06 exploited a variety of data on faint distant galaxies from the Spitzer Space Telescope IR camera IRAC at $3.6\ \mu$m and from ground-based observations in the $2.2\ \mu$m K-band.
In particular the evolution of the galaxy luminosity functions is inferred from the analysis of multi-wavelength data in the Chandra Deep Field South (CDFS) area, including very deep high-resolution imaging by HST/ACS. Their reference catalogue of faint high-redshift galaxies comes from a deep ($S_{3.6}\geq 1\ \mu$Jy) IRAC photometric catalogue. 
These imaging data in the field are complemented with extensive optical spectroscopy to obtain redshift distances, while deep K-band imaging is also used to derive further complementary statistical constraints and to assist in the source identification and SED analysis. 

A highly reliable IRAC 3.6$\mu$m sub-sample of 1478 galaxies with $S_{3.6}\geq 10\ \mu$Jy has been selected, half of which has spectroscopic redshifts, while for the remaining half highly accurate photometric redshifts were used. 
This very extensive dataset was exploited to estimate the galaxy luminosity and stellar mass functions of galaxies, and to assess evolutionary effects as a function of redshift. The luminosity/density evolution was further constrained with the number counts and redshift distributions. 

The deep ACS imaging allowed us to differentiate these evolutionary paths by morphological type,  considering that the various galaxy morphological classes experience different physical and evolutionary properties. 
With the ACS data, structural classification was found to be reliable at least up to $z\sim 1.5$ for the two main galaxy classes, the early-type (E/S0), or spheroidal, and the late-type (Sp/Irr), or starforming, galaxies.

The luminosity function data derived from the CDFS sample, as well as the estimate of the stellar mass function above $M_\ast h^2=10^{10} M_\odot$ for the spheroidal subclass, consistently showed a progressive lack of such objects starting at $z\sim 0.7$, paralleled by an increase in luminosity.
A similar trend, with a more modest decrease of the mass function, is also shared by spiral galaxies, while the irregulars/mergers show an increased incidence at higher z.
Remarkably, this decrease of the comoving density with redshift of the total population appears
to depend on galaxy mass, being stronger for moderate-mass, but almost absent until $z=1.4$ for high-mass galaxies, thus confirming previous evidence for a "downsizing" effect in galaxy formation.
The favoured interpretation of the evolutionary trends for the two galaxy categories is that of a progressive morphological transformation (due to gas exhaustion and, likely, merging) from the star-forming to the passively evolving phase, starting at $z\geq 2$ and continuing to $z\sim 0.7$. The rate of this process appears to depend on galaxy mass, being already largely settled by $z\sim 1$ for the most massive systems. Similar results were found by Bundy et al. (2005).

AF06 used simple evolutionary models to fit the fast convergence of the number counts and redshift distributions, and the evolutionary mass function.
In their approach, three main galaxy classes are considered as dominating the   galaxy populations and being potentially characterized by different evolutionary histories: spheroidal (E/S0) galaxies, quiescent spirals, and an evolving population of irregular/merger systems (hereafter the starburst population).
On the contrary, it is easily shown that active galactic nuclei do not significantly contribute to the extragalactic counts in either the optical or the near-IR, whereas they significantly contribute to the mid- and far-IR counts, as discussed below.

Synthetic spectral energy distributions (SED) have been adopted to transform the various statistics, like the galaxy luminosity functions needed to calculate the number counts and photon densities, from one wavelength to another. Different astrophysical prescriptions have been adopted to calculate the spectra of different galaxy populations. In particular, for early-type galaxies a rapid build-up ($t_{SF}=0.1~Gyr$) of the stellar content has been adopted with an initially high star-formation (SF) efficiency, with the corresponding SF law having a maximum at a galactic age of 0.3 Gyr and quickly decaying thereafter. During this rapid star-forming phase, the galaxy emission is assumed to be extinguished by $A_V=6$ magnitudes.

For late-type (spirals and irregular) and starbursting galaxies, a longer SF timescale was adopted, $t_{infall}=4~Gyr$, with a correspondingly lower initial SF efficiency. In such a case, the star-formation is a more gradual process taking place during the whole Hubble time.

Our assumed local luminosity functions (LLF) have been derived from Kochanek et al. (2001) based on a large $K$-band selected local galaxy sample. These functions have been separately calculated for both the early-type and late-type galaxy classes based on a morphological analysis. 
Transformation from 2.2 to 3.6 $\mu$m is performed with the SED templates for the two classes at the present cosmic time.

A particularly important constraint comes from the Spitzer/IRAC observations at 8 $\mu$m of the faint number counts (Fazio et al. 2004; Lonsdale et al. 2004) and the local galaxy luminosity function (Huang et al. 2007). These data very significantly constrain the photon background density at the junction between the stellar photospheric emission in the optical/near-IR and the far-IR dust emission peak (see Fig. \ref{bkg}).
Given the uncertainty in the 8 $\mu$m galaxy counts by Fazio et al. at the bright fluxes, we have rebuild them by including the observations by the SWIRE collaboration (Surace et al., in preparation; Lonsdale et al. 2003) over a large (10 sq.deg.) sky area, to minimize the effect of the local source clustering. The z=0 diffuse light at 8 $\mu$m based on these new counts is reported as a triangular datapoint at $\lambda=8\ \mu$m in Fig. \ref{bkg}. 

Our fits to the observed counts and luminosity functions in the interface region from $\lambda=12$ to 6.7 $\mu$m is obtained by summing up the two independent contributions by the far-IR emitting and optically emitting galaxies.
Outside this wavelength interval, the two separate populations of galaxies largely dominate the counts and the contribution from the other one is neglected.

\subsubsection{The spiral population}

In the AF06 scheme, once formed at a given redshift, the comoving number densities of the spiral population remains constant, while the galaxy luminosities evolve following the evolution of their stellar content. 
This choice reflects the assumption that, once having acquired its final morphological structure within the Hubble sequence, a normal spiral galaxy evolves only due to the secular change of the integrated stellar spectrum.
For the spiral galaxy class a high redshift of formation ($z_{form}=5$) was assumed.

\subsubsection{A population of fast-evolving starbursts}
\label {FBO}

There is evidence in optical and near-IR surveys of faint high-redshift galaxies of the presence of a numerous population of irregular/merging systems at high-redshifts, likely suggesting luminosity as well as density evolution back in cosmic time.  This evidence comes partly from the faint galaxy counts in the B-band (Ellis 1997), and even more from far-IR observations (e.g. Franceschini et al. 2001; Elbaz et al. 2002).
These data indicate that a large component of the faint galaxy population with irregular/merging morphologies is subject to both luminosity and number density evolution going back in cosmic time.

A population of starburst galaxies was then included in our model, whose comoving number density $\rho(z)$ evolves according to:
\begin{equation}
\rho(z) \propto \rho(z_0) \times (1+z) 
\end{equation}
for $z<1$, keeping constant above, and whose luminosities $L(z)$ also increase with redshift as
\begin{equation}
L(z) = L(z=0) \times exp[k\cdot \tau(z)] ,
\end{equation}
where $\tau(z)=1-t_H(z)/t_H(z=0)$ is the look-back time in units of the
Hubble time $t_H(z=0)$, and where the evolution constant is $k=1.7$ for $0<z<2$, and $k=const$ at $z\geq 2$  [such that $L(z=1)\simeq 2.6\cdot L(z=0)$ and $L(z=2)\simeq 4\cdot L(z=0)$]. 


\subsubsection{An empirical evolutionary scheme for spheroidal galaxies}

The model adopted for spheroidal galaxies describes a situation in which massive ellipticals mostly form at moderate redshifts ($1<z<2$) through the merging of smaller units down to relatively recent epochs. Their formation is not a single coeval process, but is rather spread in cosmic time.
This was achieved by splitting the local spheroids into seven sub-populations, each one forming at different redshifts (in the redshift interval $0.9 < z < 1.6$, see details in AF06). For simplicity, all sub-populations are assumed to have the same mass and luminosity functions and to differ only in the normalizations, whose total at $z=0$ has to reproduce the local observed luminosity function (Kochanek et al. 2001).

This model, called $Protracted$-$Assembly$ (PA), was tested against existing data on deep surveys of spheroidal (E/S0) galaxies in the near-IR bands, and was found to be consistent with all of them, particularly with the distributions of redshifts of flux-limited samples setting important constraints on their evolution.
Between two slightly different implementations of the PA formation scheme in AF06, we adopt here the model PA-1, in which stars in a sub-population of spheroidal galaxies are assumed to be coeval with the formation redshift
of the population, $z_{form}$.

Adding the contributions from the three galaxy classes, spirals, fast-evolving starbursts and PA-assembling speroidal galaxies, our model is able to very accurately reproduce the various cosmological observables at near-IR wavelengths, like the number counts down to faint flux limits in both the 2.2 $\mu$m $K$-band and the Spitzer 3.6 $\mu$m band. As such, the model provides an excellent fit to the extragalactic near-IR background intensity discussed in Sect.\ref{irbck} (see Fig.\ref{bkg}).
The model fits the redshift distributions for faint samples at 2.2 and 3.6 $\mu$m and, at the same time, is consistent with the evolutionary trends of the redshift-dependent galaxy mass function (see details in AF06).

\subsection{Modelling the evolution in the optical}
\label{optical}

The UV and optical spectral region is critical for establishing the opacity for photon-photon interactions of high-redshift sources, because the sub-TeV photon energies, at which the universe is more transparent, do preferentially interact with such background radiation.

For continuity, we have adopted here the same model as described for the near-IR part, that is the same set of source populations and evolutionary scheme. The AF06 near-IR multi-wavelength model has then been extrapolated to shorter wavelengths by means of synthetic spectral energy distributions (SED) associated with the various galaxy populations, which are described in detail in AF06 (their Sect. 7) and are summarized in our Sect. \ref{nirmod}.

\begin{figure}[!ht]
\centering
\includegraphics[angle=0,width=0.5\textwidth,height=0.4\textwidth]{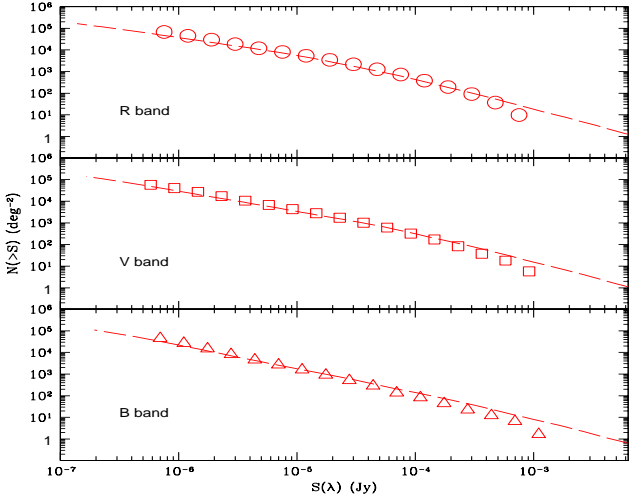}  
\caption{Comparison of the observed galaxy number counts in the B, V, R bands (shown as datapoints) with those predicted by the model in Sect. \ref{optical} (long dash lines).  Datapoints are from Berta et al. (2006).
}
\label{cint_rho}
\end{figure}

This extrapolation to short wavelengths has a limit in the fact that near-IR and optical-UV wavelengths tend to emphasize somewhat different astrophysical processes in distant galaxies. In the rest-frame near-IR, the galaxy integrated light is typically contributed by low-mass long-lived stars and to a lesser extent by high-mass stars. The optical-UV is instead more sensitive to young short-lived stellar populations. As a consequence, the optical-UV galaxy counts and background light include contributions by moderate-mass starbursting galaxies with blue optical spectra, which are relatively faint in the near-IR (the Faint Blue Object - FBO - population, see Ellis 1997). We have modelled this population with the class of fast-evolving starbursts described in Sect. \ref{FBO}: due to their substantial cosmological evolution and blue spectra, these galaxies produce increasing contributions at shorter wavelengths compared with normal spirals and spheroids, and dominate the faint optical counts.

We report in Fig. \ref{cint_rho} a comparison of our predicted with the observed optical galaxy counts (Berta et al. 2006), showing excellent agreement down to 0.65 $\mu$m (top panel, the R-band). At shorter wavelengths the faint observed counts keep slightly above the model, due to some unaccounted contribution of the FBO. The corrections to the number counts to bring them into agreement with the data are small (20\% in the V 5500\AA \ and 40\% in the 4400\AA \ UV-band). Given that the redshift distribution of FBO is essentially the same as the normal spiral and starburst galaxies, we have simply applied these small corrections to the model counts, and obtain the good fits reported in the bottom panels of Fig. \ref{cint_rho}.  
At even shorter UV wavelengths the photon number densities are very low and provide negligible contribution to the cosmic opacity, such that no corrections are needed.

\begin{figure*}[!ht]
\centering
\includegraphics[angle=0,height=0.6\textwidth,width=0.8\textwidth]{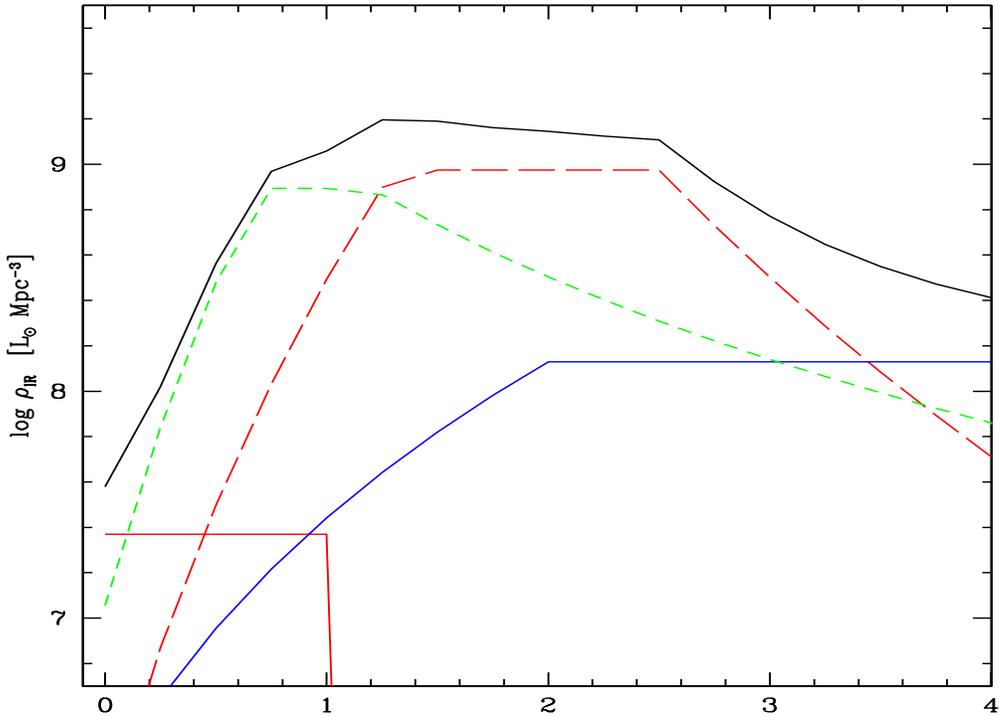}  
\caption{Evolution of the comoving bolometric luminosity density from 6 to 1000 $\mu$m for the IR-selected galaxy population, based on the model discussed in Sect. \ref{irmod}. The luminosity density is expressed here in solar luminosities per cubic Mpc.
Lower continuous red line: quiescent spiral population.
Blue continuous line: type-I AGNs.
Green short-dashed line: evolving moderate-luminosity starbursts.
Red long-dashed line: high-luminosity starbursts. 
The upper continuous line is the total emissivity. 
}
\label{rho}
\end{figure*}

\begin{figure*}[!ht]
\centering
\includegraphics[angle=0,width=\textwidth,height=0.7\textwidth]{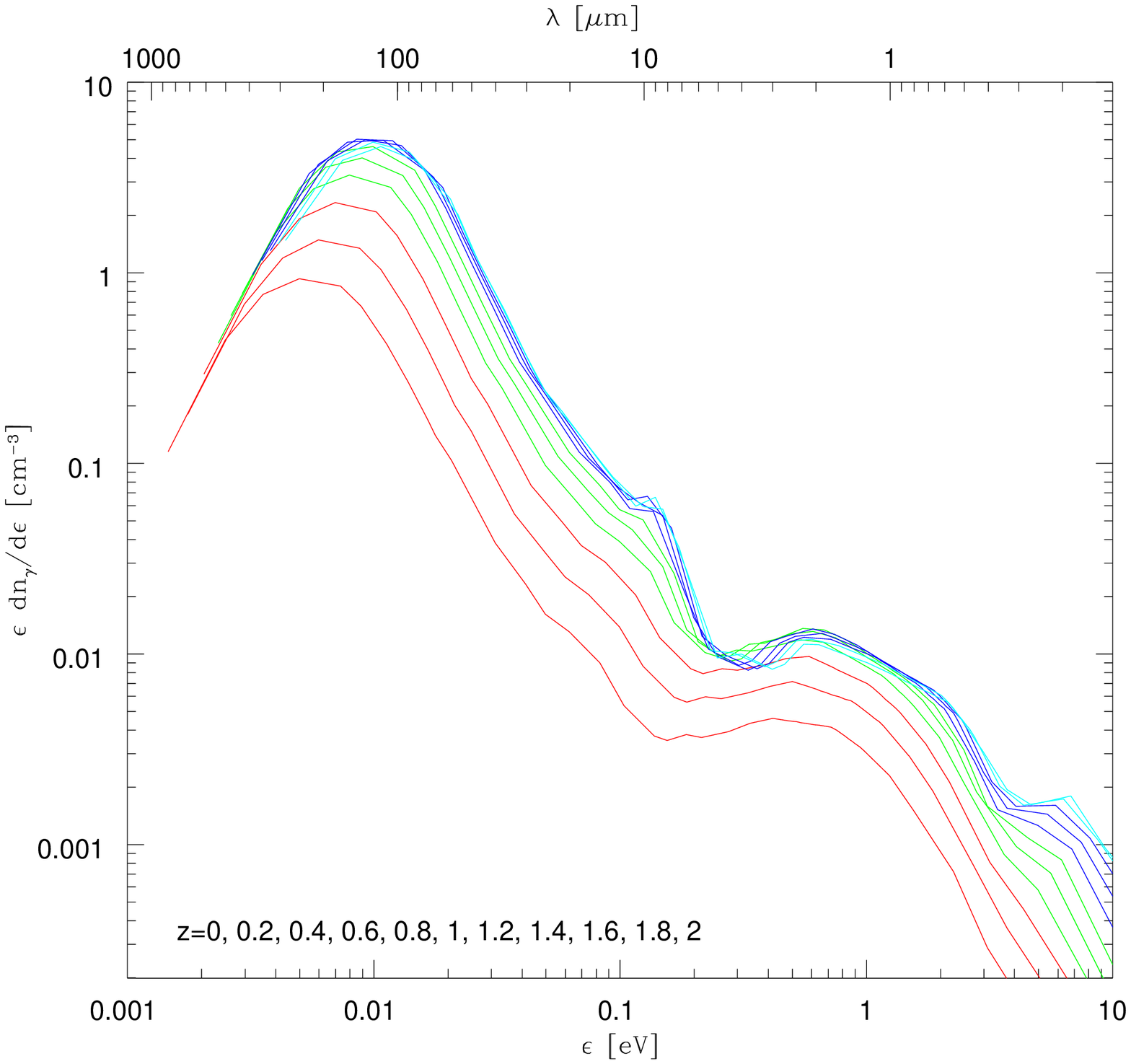}  
\caption{The redshift-dependent photon number density multiplied by the photon energy $\epsilon$. The number density here is in proper (physical, not comoving) units [$cm^3$] and $\epsilon$ in eV.
Various curves are for different redshifts: red correspond to z=0, 0.2, 0.4;  green to z=0.6, 0.8, 1.0; blue to z=1.2, 1.4, 1.6; cyan to z=1.8, 2.
Photons are generated by our model of the evolution of cosmic sources discribed in Sect. \ref{model}.
}
\label{ngamma}
\end{figure*}

\subsection{Galaxy evolution in the infrared and sub-mm}
\label{irmod}

Aiming at an accurate and detailed representation of the evolution of galaxies at long IR wavelengths, we need to develop an independent evolutionary scheme from those used to fit the optical and near-IR data. The reason for this is that the astrophysical mechanisms producing photons in the two waveband domains are completely different: while stellar photospheric emissions dominate in the latter, long wavelength photons are mostly produced by thermal emission of dust particles present in the galaxy interstellar medium (ISM). This dust absorbs the UV photons produced by young stars, is heated to a few tens of degrees and re-radiates typically between 10 and 1000 $\mu$m. Early far-IR observations of the local universe by the IRAS satellite (Soifer, Neugebauer and Houck 1987) have revealed that extragalactic sources are energetically dominated by luminous (LIRGs) and very luminous (ULIRGs) IR galaxies, a situation which was found to be rather different to that typical at UV wavelengths, where the majority of the UV emitters are moderately massive and luminous objects. This difference is a consequence of the fact that the most luminous galaxies are those characterized by the most intense star formation and that stars vigorously form inside dust-opaque media. Low-mass galaxies, instead, do not typically contain a ISM rich enough in dust to be optically-thick to dust absorption, and are faint in the far-IR (see Lagache, Puget and Dole 2005, Franceschini et al. 2001, for reviews).
Far-IR emission of galaxies is then an optimum tracer of the ongoing rate of star-formation (SFR) even in the most luminous and massive systems.

After the IRAS exploratory survey, enormous progress has been made in our understanding of the distant universe at IR wavelengths (12 to 170 $\mu$m) thanks to two dedicated cooled space observatories, \textit{ISO} and, particularly, the still operative \textit{Spitzer Space Telescope}. In spite of their primary mirror's small sizes, the two observatories have performed extensive explorations of high-redshift galaxies thanks to the extremely high sensitivity allowed by the very low background noise (Elbaz et al. 1999, 2002 and Franceschini et al. 2001 for the main ISO results;  Le Floc'h et al. 2005, Pérez-González et al. 2005,  for Spitzer results). 

Finally, the long-wavelength portion of galaxy emission, at 350 to 1200 $\mu$m, has been sampled by ground-based millimetric and sub-millimetric telescopes with sensitive bolometric cameras (SCUBA on JCMT, see e.g. Blain et al. 2002; BOLOCAM and MAMBO on IRAM, e.g. Bertoldi et al. 2003). These observations have benefited by the large photon collectors of these telescopes and by the favourable K-correction bringing the peak of galaxy dust emission at $\sim 100\ \mu$m into the mm wavebands for sources at $z$ up to 5.

With this rich observational database, to model the IR galaxy evolution we have followed a similar approach to that adopted in Sect. \ref{nirmod}, i.e. to use an empirical representation based on a backwards evolution in cosmic time of the multi-wavelength local IR galaxy luminosity functions. Here again we exploited both a direct evaluation of the evolutionary luminosity function at 24 $\mu$m and integrated observables, like the galaxy number counts at various far-IR and sub-mm wavelengths, redshift distributions for flux-limited samples, etc.
For a more complete account of this modelling see Franceschini et al. (2008).

\subsubsection{The evolutionary luminosity function of galaxies at 24 $\mu$m}
\label{ir24}

We have tested the predictions of the IR evolutionary model by comparison with the observed 24 $\mu$m rest-frame luminosity function up to $z\sim$2.5, estimated by us (Rodighiero et al. 2008, in prep.). 
This is based on a large galaxy sample combining shallow and wide-area data in common between the
SWIRE (Lonsdale et al. 2003) and the VVDS surveys (e.g. de La Torre et al. 2007), for a total of $\sim$0.85 square degrees), together with the smaller-area but deeper data available in the GOODS-South field ($\sim$130 arcmin$^2$).

The primary selection of this analysis comes from a flux-limited sample of MIPS/Spitzer 24 $\mu$m sources: in the SWIRE-VVDS area we used a bright-flux limit of $S(24\mu m)>400 \mu$Jy, which includes 1494 sources. The  GOODS-South sample reaches $S(24\mu m)>80 \mu$Jy, for a total of 614 sources.
Both samples are complete down to these limits, such that no completeness corrections were needed when computing the luminosity functions.

We have made use of the detailed multiwavelength photometric data available from HST, Spitzer and ground-based telescopes to fit the observed spectral energy distributions of each source with a suitable set of IR templates, including normal, star-forming galaxies and AGNs (Polletta et al. 2007).  These fits were used to compute accurate K-corrections for each source.
In the GOODS-South, about half of the sample has a spectroscopic redshift, and for the remaining fraction we adopted accurate photometric redshifts (for example from the MUSIC sample, Grazian et al. 2006) .
In the SWIRE-VVDS area, the spectroscopic follow-up covers only a small fraction of the field and the fraction of sources with spectroscopic information is only $\sim10$\% at the flux limit considered. However, here we benefit from the excellent quality of the photometric redshifts from Ilbert et al. (2006).

A detailed analysis of the source confusion level was performed in the GOODS-South sample, where the blending of sources at 24 $\mu$m wavelengths can affect about 20\% of the sample at the level of 80$\mu$Jy (Rodighiero et al. 2006). We have applied a robust maximum likelihood estimator to associate the optical and near-IR counterparts with the 24 $\mu$m selected sources.

The 24 $\mu$m rest-frame luminosity function has been computed up to $z\sim$2.5, and split in various redshift bins using the standard $1/V_{max}$ estimator. Our analysis shows a very good agreement with the model described in the next section, at all redshifts, implying that the observed evolution is consistent with that assumed by our evolutionary scheme.

\subsubsection{Additional statistical observables in the infrared}

We have made use of a variety of additional statistical observables to constrain our multi-wavelength evolution model of galaxies. In particular, our reference local LF was derived from complete galaxy and AGN catalogues from the IRAS 12 and 60 $\mu$m survey, as detailed in Franceschini et al. (2001). Differential number counts at 15 $\mu$m were taken from the ISO surveys, whereas deep number counts at 24, 70, and 160 $\mu$m come from the Spitzer Space Telescope. At the longest wavelengths, reliable number counts have been obtained from SCUBA observations at 850 $\mu$m (e.g. Coppin et al. 2006).

Essential constraints on the time evolution of galaxy emissivity have been inferred by us from redshift distributions of complete flux-limited samples at 15 and 24 $\mu$m (see for more details Franceschini et al. 2008).

\subsubsection{Multi-wavelength modelling of IR galaxy evolution}

Our adopted evolutionary scheme is based on the analysis by Franceschini et al. (2001), which was calibrated on the ISO database. The IR source population is modelled (similarly to Sect. \ref{nirmod}) with the contribution of four population components characterized by different physical and evolutionary properties: the first two are non-evolving normal spirals dominating the 12 $\mu$m LLF at low luminosities, and a fast evolving population of moderate luminosity objects, including starburst galaxies and type-II AGNs, with the assumption that both classes have the same evolutionary properties and that the IR spectrum is dominated by starburst emission.
A third considered component is type-I AGNs, which is modelled based on optical and X-ray
quasar surveys, and assumed to evolve in luminosity as 
$L(z)=L(0)\times (1+z)^3$ up to $z=1.5$ and $L(z)=L(0)\times 2.5^3$ above.

The local fraction of the evolving moderately-luminous starburst population is assumed to be $\sim 10$ percent of the total, roughly consistent with the local observed fraction of interacting galaxies.
The observed upturns in the 15, 24, 160 and 850 $\mu$m counts then require a strong increase with redshift of the average emissivity of the evolving population to match the peak in the  normalized differential counts, an increase which is obtained with both luminosity and number density evolution:
\[   \rho(L_{12}[z],z) = \rho_0(L_{12})\times (1+z)^{4} \ \  \ \ \ z<z_{break}  \]
\[  \rho(L_{12}[z],z) = \rho_0(L_{12})\times (1+z_{break})^{4} \ \  \ \ \ z_{break}<z<z_{max}  \]
\[ L_{12}(z) = L_{12}\times (1+z)^{3.8}   \ \  \ \ \ z<z_{break}  \]
 \begin{equation}
 L_{12}(z) = L_{12}\times (1+z_{break})^{3.8}   \ \  \ \ \ z_{break}<z<z_{max}
\label{solu}
\end{equation}
with $z_{break}=0.8$ and $z_{max}=3.7$. 

An important addition to the Franceschini et al. (2001) scheme was to introduce a fourth evolutionary component of very luminous IR objects dominating the cosmic IR emissivity at $z\simeq 2$. This latter emerges from the analysis of the galaxy samples selected by the Spitzer/MIPS deep 24 $\mu$m surveys, revealing a population of luminous IR starbursts (ULIRGs) at high-redshift, essentially absent or very rare locally, hence caracterized by an extremely fast evolution in cosmic time. Millimeter observations with SCUBA have also found clear evidence of such a very luminous galaxy population at high redshifts. The current interpretation identifies these objects with the progenitors of the spheroidal galaxies discussed in Sect. \ref{nirmod}.

The redshift evolution of the comoving volume emissivities from 6 to 1500 $\mu$m of the various populations of IR sources is reported Fig. \ref{rho}.  It is evident from the figure that the two main populations of moderate- and high-luminosity starbursts evolve in a significantly different pattern, the more luminous ones dominating at the highest redshifts.

\begin{figure}[!ht]
\centering
\includegraphics[angle=0,width=0.5\textwidth,height=0.5\textwidth]{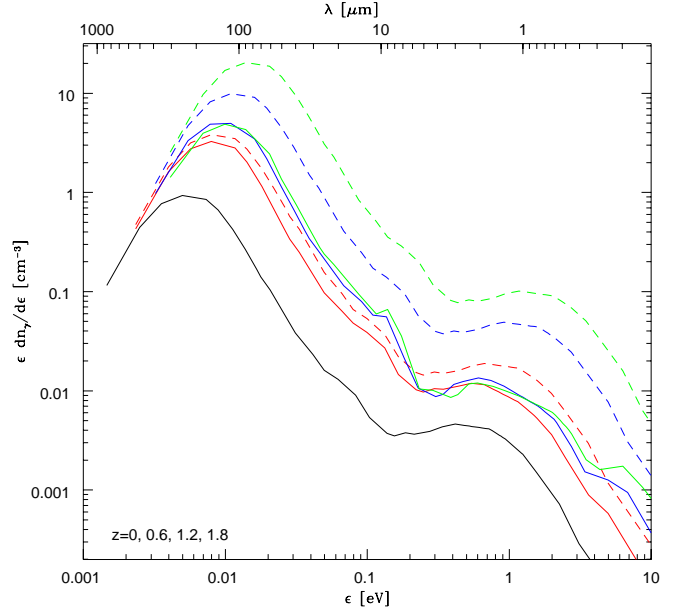}  
\caption{A comparison of the redshift dependence of the proper photon number density obtained by our galaxy evolutionary models (continuous lines) with that corresponding to the case of a non-evolving population (dashed lines). 
The lines with black, red, blue and green colors correspond to z=0, 0.6, 1.2, 1.8.
Our accounting of the evolution effects makes the predicted background photon density much lower at high-z (z$>0.6$) than the no-evolution prediction, particularly in the optical.
}
\label{ngamma1}
\end{figure}

A detailed comparison of our model of IR galaxy evolution with the data on the 24 $\mu$m luminosity functions in redshift bins discribed in Sect. \ref{ir24} shows excellent agreement up to the maximum redshift of $z=2.5$, with a total reduced $\chi^2\simeq 1.1$. The model fit is formally worse in the lowest bin at ($0<z<0.25$), due to unaccounted effects of the local large-scale structure in the luminosity function data. However, the low-z behaviour of the model has been calibrated on a variety of number count data points at bright fluxes (including the IRAS all-sky far-IR counts and those from the Spitzer SWIRE project), which guarantees excellent control of the local IR universal emissivity.

Thus our multi-wavelength analysis, while not providing a definitive explenation of the physical processes involved, gives the best possible fit to a wide variety of data on the photon emissivity of cosmic sources from the local to the distant universe.

\section{The evolutionary photon number density and the cosmic photon-photon opacity}

\subsection{The photon number density}
\label{density}

The photon number density and its evolution with cosmic time is easily calculated with the model in the previous Section. We need to compute the photonic background intensity for an observer at any redshift $z^\ast$. To do so, let us first consider the local background intensity $I_{\nu_0}(z^\ast)$ from all sources radiating between $z=z^\ast$ and the maximum redshift of the source distribution, $z_{max}$ (sources with $z<z^\ast$ do not obviously contribute to the intensity at $z^\ast$). Their contribution to the local background intensity is given by (e.g. in Peacock 2000):
   \begin{equation}
   \begin{array}{l}
I_{\nu_0}(z^\ast) = \frac{1}{4\pi} \frac{c}{H_0} \int_{z^\ast}^{z_{max}} dz\ \frac{j[\nu_0(1+z),z]}{1+z} \times \\ \\
\times [(1+z)^2(1+\Omega_m z)-z(2+z)\Omega_\Lambda]^{-1/2} ,
   \end{array}
   \label{4}
   \end{equation}
for the standard case of a flat universe with $\Omega_m + \Omega_\Lambda=1$. Here $j[\nu_0]$ is the galaxy comoving volume emissivity at redshift $z$:
 \begin{equation}
 j[\nu_0,z] = \int _{L_{min}} ^{L^{max}} d\log L_{\nu_0} \ \cdot n_c (L_{\nu_0},z) \ \cdot K(L_{\nu_0},z) \ \cdot L_{\nu_0} ,
 \end{equation}
where $K(L,z)$ is the K-correction
 \begin{equation}
 K(L,z) = (1+z) \cdot \frac{L_{\nu_0(1+z)}} {L_{\nu_0}}  ,
    \label{Kc}
 \end{equation}
and $n_c$ is the comoving luminosity function at the redshift $z$ expressed as the number of galaxies per $Mpc^3$ per unit logaritmic interval of the luminosity $L$ at frequency $\nu_0$.
 Note that the $[1+z]^{-1}$ factor in the first row of eq. \ref{4} assumes that $j$ and $n_c$ are per comoving volume; the use of proper volumes would require a factor $[1+z]^{-4}$. 
The local background intensity reported in Fig. \ref{bkg} (continuous line) is calculated from eq. \ref{4} from our evolution model by setting $z^\ast=0$. 

Moving now to the units in the proper volume, from eq. \ref{4} the specific radiant energy density [e.g. in $erg/sec/cm^3/Hz$] in the proper volume at the redshift $z^\ast$ is easily calculated as:
 \begin{equation}
    \begin{array}{l}
 \rho(z^\ast, \nu_0) = \frac{4\pi}{c} (1+z^\ast)^3 \cdot I_{\nu_0}(z^\ast) \ =  \\ \\
 = \frac{1}{H_0} (1+z^\ast)^3 \cdot \int_{z^\ast}^{z_{max}} \frac{dz \cdot j[\nu_0(1+z),z]}
{ (1+z)\  [(1+z)^2(1+\Omega_m z)-z(2+z)\Omega_\Lambda]^{1/2} } ,
    \end{array}
 \end{equation}
where the factor $(1+z^\ast)^3$ corrects the energy density from local to $z^\ast$. As discussed in Peacock (2000), this spectral energy density is only modified by the change in photon density, because the increase $\propto(1+z)$ of the photon energy $\epsilon$ is exactly compensated by the increased bandwidth $d\epsilon$.
Finally, the photon proper number density is:
\begin{equation}
 \frac{dn_\gamma(\epsilon, z^\ast)}{d\epsilon} =  \rho(z^\ast, \nu_0) \frac{1}{\epsilon} ,
 \end{equation}
where $\epsilon = h\nu_0$ is the photon energy.  A plot of our model predictions for the differential photon proper number density (in proper, not comoving, volume units) appears in Fig. \ref{ngamma}, while the corresponding data are reported in Tables \ref{tngamma} and  \ref{tngamma1}. 
The two galaxy emission peaks, due to photospheric stellar emission (rest wavelength 1 $\mu$m) and dust re-radiation emission (rest wavelength 100 $\mu$m), are clearly apparent in the figure at all redshifts at photon energies of $\epsilon\simeq 1$ and 0.01 eV. 

The general behaviour of the spectral densities is that of an increase of the photon \textit{proper} density with redshift due to the Hubble expansion. However, such an increase is not simply proportional to the volume factor $(1+z)^3$ for two reasons. The first one is that photons are progressively generated by galaxy populations during most of the Hubble time to the present, so that the photon \textit{comoving} number density decreases with redshift. The second effect is due to the K-correction (eq. \ref{Kc}), which tends to accumulate photons as a function of time at energies where $K(L,z)$ is larger. Due to the K-correction effect, the evolution of the proper density $n_\gamma$ is minimal in the submm ($\epsilon<0.01\ eV$) and in the near-IR ($\epsilon<0.8\ eV$), while it is maximal at $0.01<\epsilon<0.1\ eV$.  
Strong evolution is also apparent at high photon energies ($\epsilon>4\ eV$, $\lambda<0.4\ \mu$m) in Fig. \ref{ngamma}, due to the fast evolution of the UV emissivity between z=0 and z=0.5-1 (appearence of the FBO population), as explained in Sects. \ref{nirmod} and \ref{optical}.

The effects of the different rates of cosmological evolution of the background sources are also evident in Fig. \ref{ngamma}. The evolution of galaxy population emissivity is greater in the IR ($\epsilon<0.2\ eV$) than in the optical/near-IR ($\epsilon>0.2\ eV$), because photons are produced at larger redshifts in the IR and lower redshifts in the optical. This reflects the fact that there is a larger increase in the proper photon density with $z$ in the IR and a lower one in the optical/near-IR. In the latter case, the density increase due to the expansion is almost compensated by the quick decrease with increasing $z$ of the number of photons available (most of them are produced at low $z$).

Minor bumps at $\epsilon \simeq$ 0.15, 2, and 7 eV in the spectra for $z=1$ to 2 correspond to the IR emission features by polycyclic aromatic hydrocarbons (PAH), the Balmer jump (4000 \AA ) and the UV emission by young stars, respectively. These are more easily observed at high-z because the smearing effect of the redshift is less there, while it is more substantial at lower z, where photon spectra from a wide redshift range are cumulated.

A further illustration of the effects of cosmological evolution is given in Fig. \ref{ngamma1}, where  
the redshift dependence of the proper photon number density obtained by our galaxy evolutionary models (continuous lines) is compared with that corresponding to the case of a non-evolving population (dotted lines). As we can see there, our detailed accounting for the evolution effects makes the predicted background photon density to be much lower at high-z (z$>$0.6) than the no-evolution prediction. The differences between the evolutionary and the no-evolution results are particularly evident in the optical,  more than they are in the far-IR. This is due to the fact that the optical and near-IR backgrounds are produced at lower redshifts on average than those at longer wavelengths, hence their comoving density decreases faster with $z$.

The results in Fig. \ref{ngamma}, although sharing similar behavior, also show significant quantitative differences to those reported by Stecker et al. (2006) and Stanev et al. (2006). This happens particularly between $\epsilon \sim$ 0.1 to 10 eV, likely due to the very different treatment of the evolution of galaxy and photon background in the optical/near-IR. This has a significant impact on the analysis of BLAZAR spectra from the GeV to the TeV energies.

\begin{figure*}[!ht]
\centering
\includegraphics[angle=0,width=0.8\textwidth,height=0.5\textwidth]{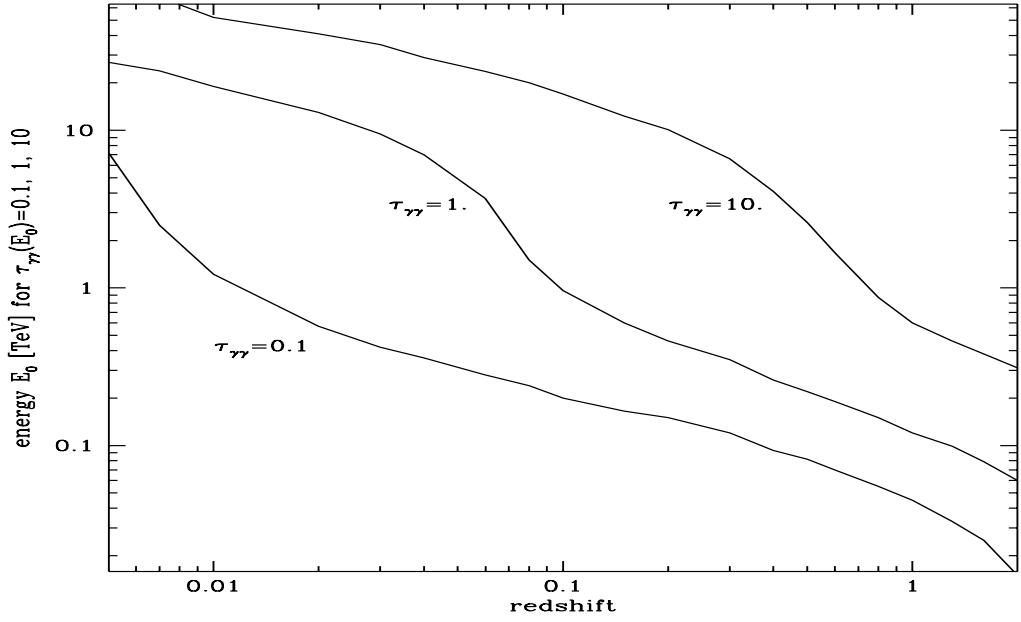}  
\caption{
The energies corresponding to optical depth values of $\tau=0.1$, 1 and 10 for photon-photon collisions, as a function of the redshift distance of the source.
}
\label{taufig}
\end{figure*}

\begin{figure*}[!ht]
\centering
\includegraphics[angle=0,width=0.8\textwidth,height=0.6\textwidth]{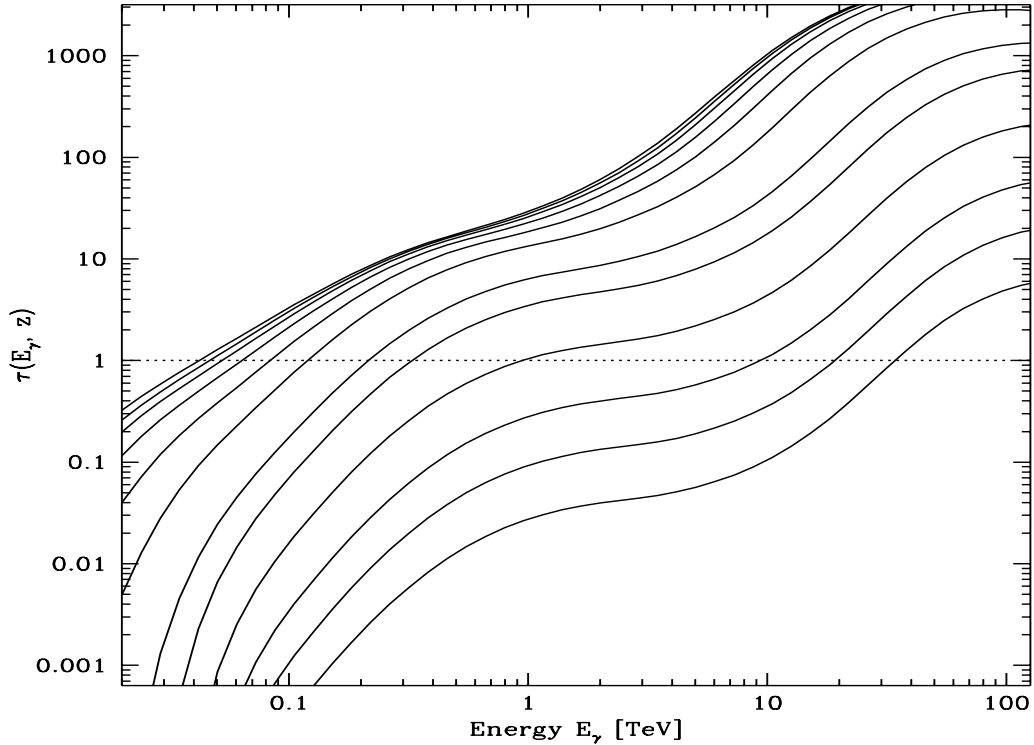}  
\caption{
The optical depth by photon-photon collision as a function of the photon energy for sources located at $z=0.003$, 0.01, 0.03, 0.1, 0.3, 0.5, 1, 1.5, 2, 2.5, 3, 4 from bottom to top.
}
\label{tauE}
\end{figure*}

\subsection{The cosmic photon-photon opacity}
\label{opacity}

Once the redshift-dependent background photon density is known, it is straightforward to compute the cosmic opacity for photon-photon interactions from the pair production cross-section (Heitler 1960):
\begin{equation}
   \begin{array}{l}
 \sigma_{\gamma\gamma} (E_\gamma,\epsilon,\theta) = \frac{3\sigma_T}{16} \cdot (1-\beta^2) \\ \\
\times \left[2\beta(\beta^2-2) + (3-\beta^4) \ln \left(\frac{1+\beta}{1-\beta}\right)\right] , \ 
   \label{sigma}
   \end{array}
   \end{equation}
where $\epsilon$ is the energy of the background photon, $E_\gamma$ that of the high-energy colliding one, $\sigma_T$ the Thompson cross-section, and where the argument $\beta$ should be computed as:
\begin{equation}
\beta \equiv (1-4m_e^2c^4/s)^{1/2};  \ \ \ s \equiv 2 E_\gamma \epsilon x ; \ \ \  x\equiv (1-\cos \theta) , 
  \label{beta}
\end{equation}
$\theta$ being the angle between the colliding photons. 
This cross-section implies that the absorption is maximum for photon energies
\[
\epsilon_{max} \simeq 2 (m_e c^2)^2/E_\gamma \ \simeq 0.5 \left(\frac{1\ TeV}{E_\gamma}\right)\ eV ,
\]
or, in terms of photon wavelength,
\begin{equation}
\lambda_{max} \simeq 1.24 (E_\gamma [TeV])\; \mu m .
\label{energy}
\end{equation}
The optical depth for a high-energy photon $E_\gamma$ travelling through a cosmic medium filled with low-energy photons with density $n_\gamma (z)$ from a source at $z_e$ to an observer at the present time is:
\begin{equation}
\tau(E_\gamma,z_e)  = c\int_0^{z_e} dz {dt \over dz } \int_0^2 dx {x \over 2} \int_{\frac{2 m_e^2c^4}{E_\gamma \epsilon x (1+z)}}^\infty d\epsilon
 \frac{dn_\gamma(\epsilon, z^\ast)}{d\epsilon} \sigma_{\gamma\gamma} (\beta)
\label{tau}
\end{equation}
where $\sigma_{\gamma\gamma}$ is given by eq. \ref{sigma}, and where the factor $s$ in eq. \ref{beta} should be replaced by
$$s \equiv 2 E_\gamma \epsilon x (1+z) . $$
For a flat universe, the differential of time to be used in eq. \ref{tau} is:
\[ dt/dz = \frac{1}{H_0 (1+z)} \left[(1+z)^2 (1+\Omega_m z) - z(z+2)\Omega_\Lambda \right]^{-1/2} . \]

We report in Fig. \ref{taufig} the energy $E_\gamma$ corresponding to optical depth values for photon-photon collisions with $\tau(E_\gamma)=0.1$, 1 and 10, as a function of the redshift distance of the source. We have not considered here the effects of the absorption by interaction with the CMB photons because it is always negligible in the parameter range of interest to us.

We also report in Fig. \ref{tauE} and in Table \ref{tautab} the optical depth as a function of energy for sources located at different redshifts. Detailed predictions for the photon-photon optical depths based on our modelling will be kept updated and publicly distributed on the WEB site
http://www.astro.unipd.it/background .

\begin{figure*}
\centering
\begin{minipage}{0.35\textheight}
\rotatebox{0}{\resizebox{9.cm}{!}{
\includegraphics{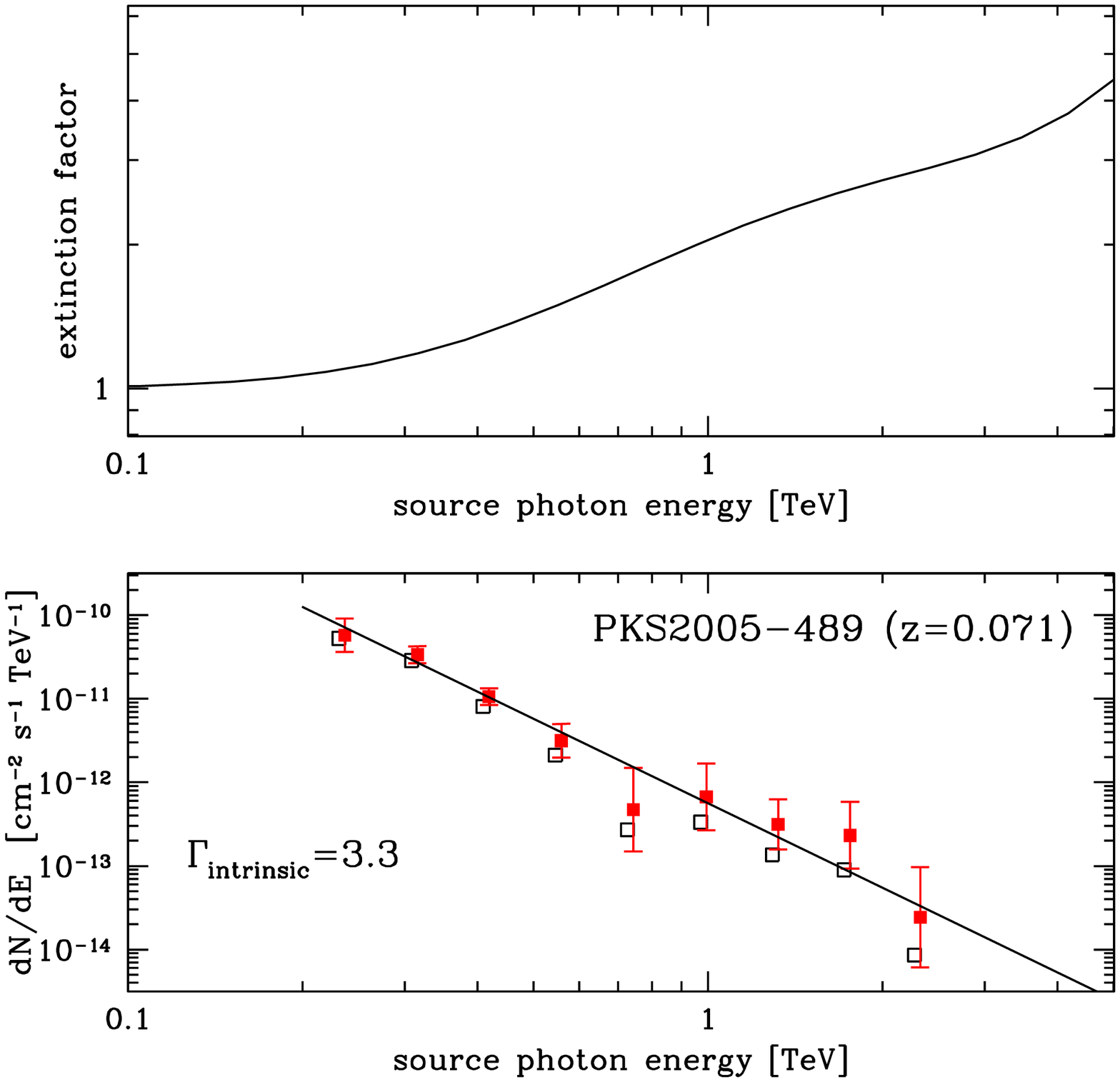}}} 
\end{minipage}
\begin{minipage}{0.35\textheight}
\resizebox{9.cm}{!}{
\includegraphics{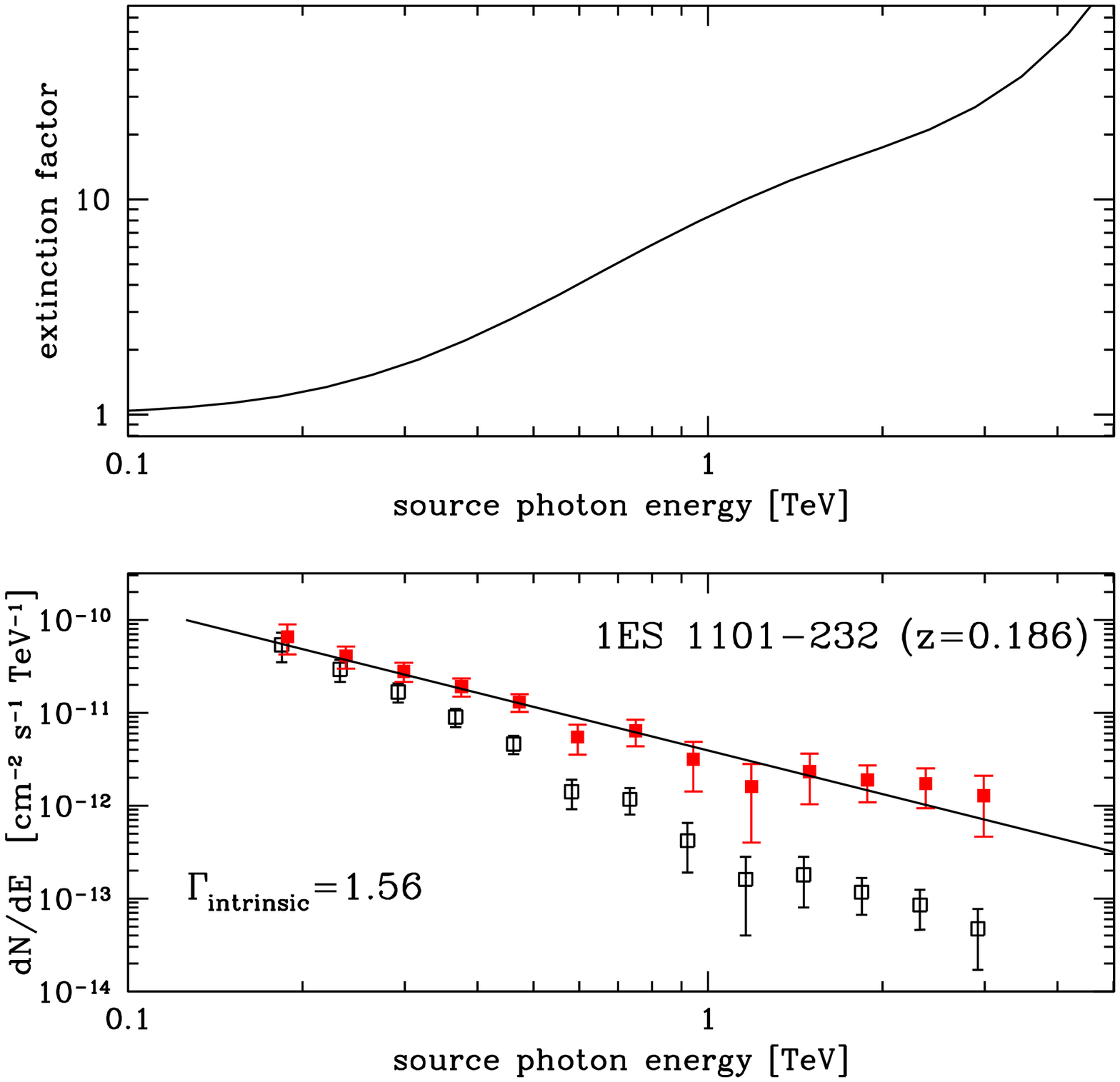}}
\end{minipage}
\caption{\textit{Top left: } Absorption correction estimated for the spectrum of the source PKS 2005-489 (z=0.071) (Aharonian et al. 2005).
\textit{Bottom left:} the observed (open black) and absorption-corrected (filled red) spectrum.
\textit{Top right: } absorption correction for the source 1ES 1101-232 at $z=0.186$ (Aharonian et al. 2006).
\textit{Bottom right:} the observed (open black) and absorption-corrected (filled red) spectrum.   
Note the slight energy shift of the corrected spectrum that we applied for clarity. } 
\label{1es1101}
\end{figure*}



%

\begin{figure*}
\centering
\begin{minipage}{0.35\textheight}
\rotatebox{0}{\resizebox{9.cm}{!}{
\includegraphics{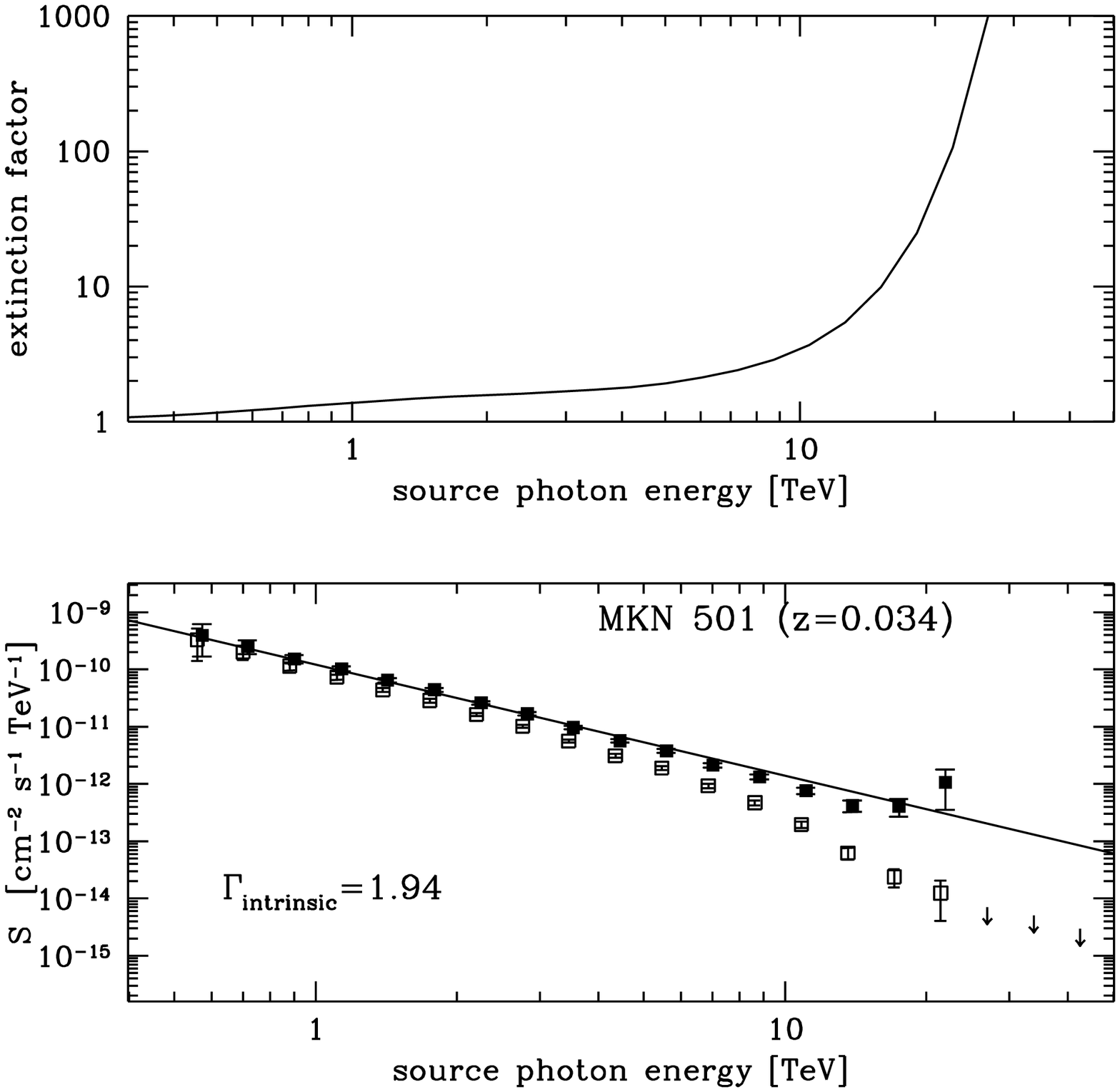}}} 
\end{minipage}
\begin{minipage}{0.35\textheight}
\resizebox{9.cm}{!}{
\includegraphics{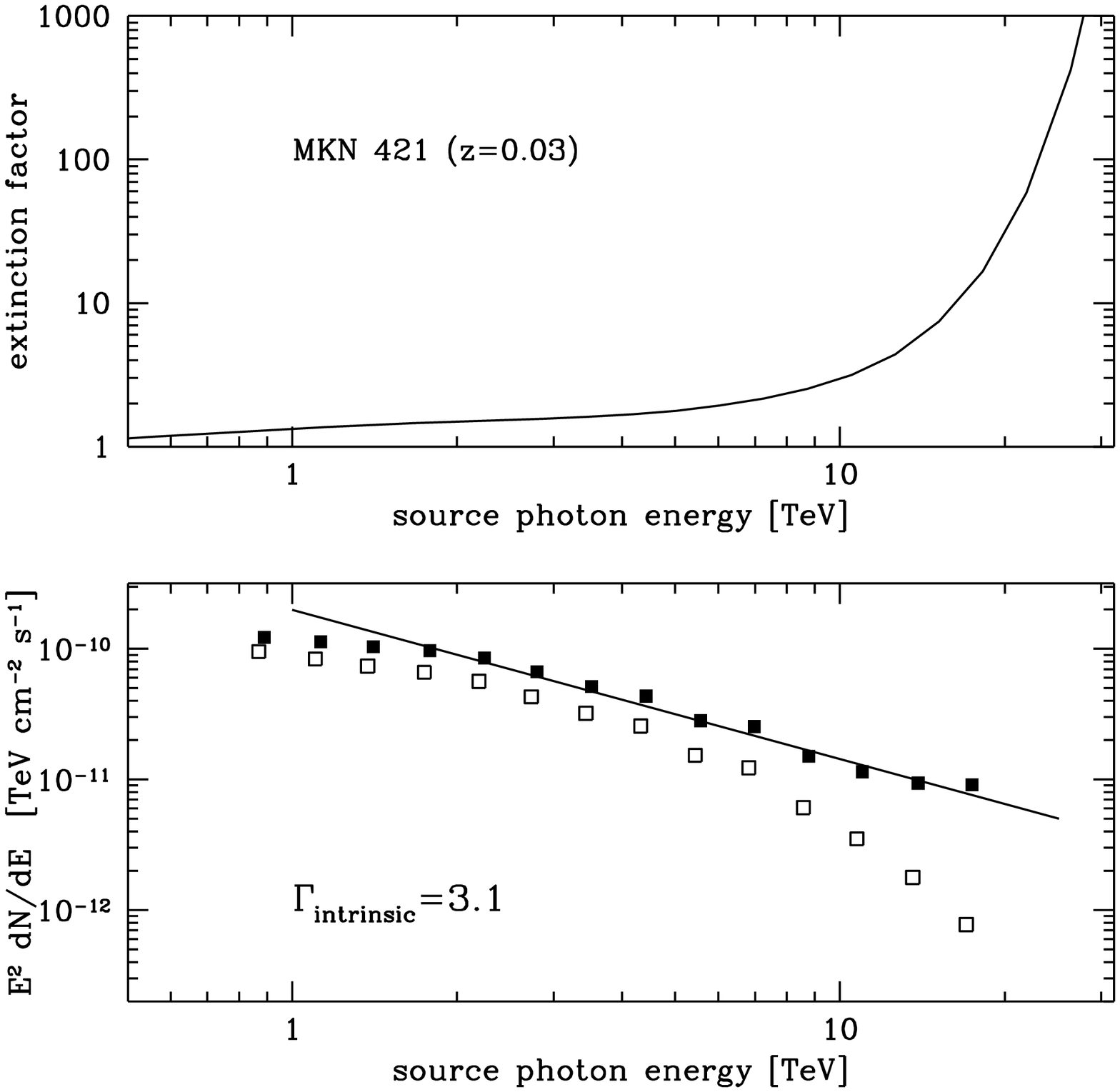}}
\end{minipage}
\caption{\textit{Top left: } Absorption correction for the source MKN 501 at $z=0.034$ (Aharonian et al. 1999).
\textit{Bottom left:} the observed (open black) and absorption-corrected (filled red) spectrum.
\textit{Top right: } absorption correction for the source MKN 421 at $z=0.03$ (Aharonian et al. 2002).
\textit{Bottom right:} The observed (open black) and absorption-corrected (filled red) spectrum.  } 
\label{mkn421}
\end{figure*}


\section{Applications to high-energy observations of distant BLAZARs}

\subsection{Effects of the minimal background from sources}

Our improved modelling of the cosmic background photon density, its evolution with time, and the cosmic photon-photon opacity have been tested by us with the analysis of the high-energy spectra of a few BLAZARs detected at TeV energies. We consider here photon backgrounds produced by discrete cosmic sources as discussed in the previous sections and exclude any purely diffuse component in a first application. The effects of the latter are discussed in Sect. \ref{PIII} below.

We have selected a small representative sample of four low-redshift BLAZARs with good-quality spectro-photometric data from Cherenkov telescopes. Two of them are the classical local objects MKN 421 and MKN 501, which have been the targets of continuous observations since the early 1990's. The other two are more distant sources, PKS2005-489 (z=0.071) and 1ES 1101-232 (z=0.186), this latter being the most distant BLAZAR for which a good TeV spectrum has been published (Aharonian et al. 2006).

Eq. \ref{energy} synthezises the relationship between the energies of background photons and those of high-energy interacting photons. Gamma-rays with $E_\gamma<1\ TeV$ interact preferentially with background photons in the optical at $\lambda\leq 1\ \mu$m, whereas the far-IR background at $\lambda>20\ \mu$m absorbs the highest-detectable energy gamma-rays at $E_\gamma>10\ TeV$.
Due to their faintness, the two higher-redshift objects, whose spectra are reported in Fig. \ref{1es1101}, have been detected only at moderate energies ($E_\gamma<3\ TeV$) by the HESS observatory, so that their emitted high-energy photons interact exclusively with the optical/near-IR background. The results of our analysis reported in Fig. \ref{1es1101} indicate that our opacity corrections produce source rest-frame spectra consistent with power laws with relatively steep photon indices. In particular, for PKS 2005-489 the photon index decreases, after the opacity correction, from the observed value of $\Gamma_{intrinsic}=4$ reported by Aharonian et al. (2005) to the steep source-intrinsic value of $\Gamma_{intrinsic}=3.3$. For the more distant object 1ES 1101-232, the larger correction results in $\Gamma_{intrinsic}\simeq 1.6$ at the source, still close to the canonical limiting value of $\Gamma=2$.

Even for 1ES 1101-232, the redshift is not so high to imply relevant corrections due to the evolution of the background photon density, as illustrated in Fig. \ref{ngamma}. However, 
the detection of TeV sources at much higher redshifts (see e.g. the results on the z=0.54 BLAZAR 3C279 with the MAGIC telescope by Teshima et al. 2007) opens the field for testing the evolving background against the observations soon.

To probe the interaction of high-energy gamma-rays with the far-IR background photons, hence constraining the far-IR branch at $\lambda> 10\ \mu$m in Fig. \ref{bkg}, one would need to observe the highest-energy tail of the source TeV spectra at $E_\gamma \geq 10\ TeV$. Due to the limited sensitivity of current Cherenkov telescopes and the tiny flux at such extreme photon energies, this has been possible only for the closest objects reported in Fig. \ref{mkn421}. The results of the photon-photon opacity corrections for these two objects show intrinsic spectral shapes again consistent with canonical values ($\Gamma_{intrinsic}\simeq 2$ and 3 for MKN 501 and MKN 421, respectively). In these cases, however, the uncertainties become significant in the high energy tail and the corrected spectrum, particularly that of MKN 501, may turn out to be slightly over-corrected (see a tendency for an upturn at the highest energies), which would imply an excess background photon density in the far-IR for our model. This event however appears rather unlikely, given that the gamma-ray photons at $E_\gamma\sim 10\ TeV$ interact with background photons at 15 to 24 $\mu$m, exactly those accurately probed by the two dedicated IR observatories, ISO and Spitzer. 
The 24 $\mu$m surveys with the Spitzer MIPS camera are particularly deep and photometrically  well calibrated (Sect. \ref{irmod}), increasing our confidence that our estimated background at those wavelengths is a robust lower limit.

All tests that we have made have shown that our detailed account of the background radiation and its evolution with cosmic time entails pair-creation opacities and source spectra that, once corrected for these opacity terms, are consistent with our current knowledge about the generation of photons in BLAZARs at frequencies $\nu > 10^{24}\ Hz$ (see e.g. Aharonian 2001 and Katarzynski et al. 2006 for reviews). Under standard conditions for particle acceleration (either leptonic, hadronic, or shock acceleration models), the intrinsic spectra are expected to be softer than $\Gamma_{intrinsic}\simeq 2$, while only in rather extreme cases is the photon index as hard as 1.5 (Aharonian et al. 2006). Other constraints come from the fact that TeV spectra are expected to be typically convex, following the shape of the lower-energy optical/X-ray synchrotron spectrum: drastic upturns with energy in TeV spectra of distant $z>0.1$ objects, although not completely excluded, seem very unlikely and also not observed in any of the local objects (e.g. MKN 421 and 501) well-studied up to $E_\gamma>10 \ TeV$.

\begin{figure}[!ht]
\centering
\includegraphics[angle=0,width=0.5\textwidth,height=0.65\textwidth]{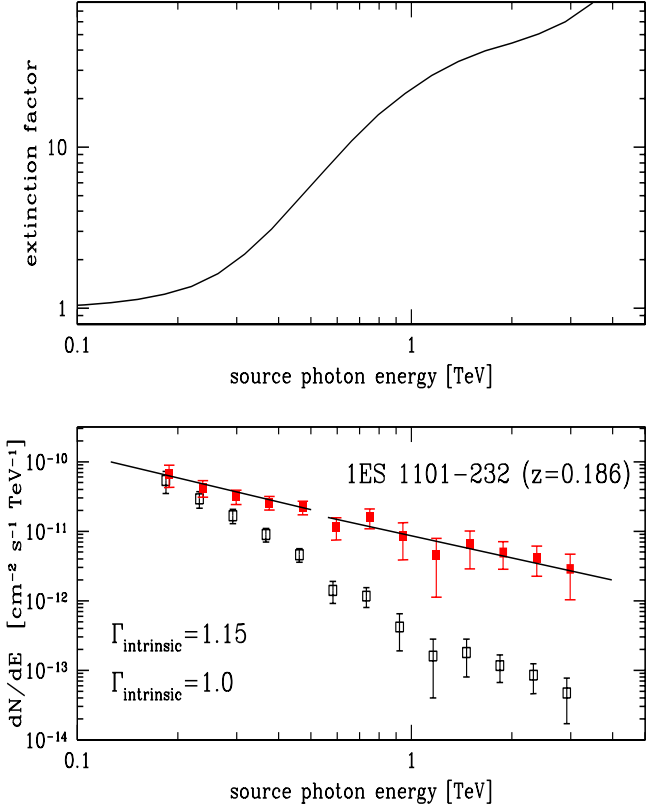}
\caption{
The effect on the spectrum of the source 1ES 1101-232 of the inclusion of a truly diffuse background in addition to that from resolved sources. The excess background is assumed to have the same spectrum as estimated by Matsumoto et al. (2005), but is down-scaled by a factor of 5. The corresponding total spectral intensity at z=0 is resported as a thick dashed line in Fig. \ref{bkg}.      
  \textit{Top:} spectral correction for $\gamma-\gamma$ opacity.
\textit{Bottom:} The observed (open black) and absorption-corrected (filled red) spectrum, the two best-fit values of $\Gamma_{intrinsic}=1.15$ and 1 refer to the lower and higher energy portions of the spectrum.
} 
\label{1ES1101PIII}
\end{figure}

\subsection{Constraints on truly diffuse backgrounds from unresolved sources}
\label{PIII}

An important aspect of our analysis is that, over the whole UV-optical-IR range, there seems to be very little room for truly diffuse background components in addition to the integrated light of galaxy populations treated in the present paper.
Such diffuse backgrounds likely contribute moderate additional flux to that originating from resolved sources, to prevent the cosmic opacity being even larger than our minimal estimated value.
Evidence for such diffuse excess light has been reported from COBE and IRTS observations at the level of several tens of nW/m$^2$/sr, that is some factors of 3 to 6 above the integrated galaxy light (see Sect. \ref{irbck}).
This question has been a subject of lively debate for several years, due to its implications.

The most interesting recently considered source of the excess near-IR background is that possibly originating from highly redshifted ($z>10$) zero-metal Population III stars. Kashlinsky et al. (2004) argue that the excess flux $I_{excess}$  in eq. \ref{FPIII} indicated by the IRTS observations of Matsumoto et al. (2005) may be explained in principle by the conversion of a fraction $f_\ast$ of all baryons into very massive stars producing photons with an efficiency $\epsilon$. Assuming for the total baryon density the value indicated by WMAP, $ \rho_b = { \Omega_b \ 3H_0^2 (1+z)^3 / 8 \pi G } $, 
with $\Omega_b\simeq 0.044\ h^{-2}$, this Population III stellar activity would produce a local photon intensity of (e.g. Franceschini et al. 2001):
\begin{equation}
 I_{excess} \simeq 12.3 \ 10^{-9} {\Omega_b h^2 \over 0.044} {f_\ast \over 0.01} \left(10 \over 1+z_\ast\right) {\epsilon \over 0.007}  \ W/m^2/sr .
\end{equation}
This, compared with the value indicated in eq. \ref{FPIII}, would require that $f_\ast\sim$3\% of the whole baryons would be converted into stars at $z\sim 10$ and would radiate with maximal efficiency to produce the excess near-IR background (Kashlinsky 2006).

As argued in Aharonian et al. (2006), all this is strongly limited by the the TeV observations of high-z BLAZARs. We report in Fig. \ref{1ES1101PIII} the analysis of the TeV spectrum of the z=0.186 source 1ES 1101-232, in which the opacity term for photon-photon collision is evaluated considering both the background photons from resolved sources and the diffuse ones of primeval origin. To the resolved-source background we then added a truly diffuse component with the same spectral shape as in Matsumoto et al. (2005), but variable normalization. For the latter component we assumed a redshift evolution in the proper photon number density given by the pure expansion factor $\frac{dn_\gamma(\epsilon_0, z^\ast)}{d\epsilon} \propto (1+z)^3$.
Increasing the normalization factor of the diffuse photon background produces a progressive flattening (hardening) of the spectrum in Fig. \ref{1ES1101PIII}, in which the higher energy portion becomes flatter than the lower energy one, contrary to expectation. 

The situation illustrated in Fig. \ref{1ES1101PIII} corresponds to a limiting case in which the two spectral branches show a hardening of the spectrum with photon energy and, at the same time, a global TeV photon index as flat as $\Gamma_{intrinsic}\sim 1$  (compared to the global value of 1.6 previously obtained based on the resolved-source background only). Both these features indicate that the corresponding opacity corrections are beyond what is allowed by physical considerations on the intrinsic spectrum, and may be considered as a reference upper limit. The latter is obtained with a truly diffuse background normalization of 20\% of the value by Matsumoto et al. in eq. \ref{FPIII}, or $I_{excess} \sim 6$  nW/m$^2$/sr between 1 and 4 $\mu$m.  
The corresponding total spectral intensity at z=0 is resported as a thick dashed line in Fig. \ref{bkg}.   

In terms of a maximally efficient Population III stellar generation with the same parameters as previously adopted, this would limit to less than 0.5\% the fraction of transformed baryons in this event. Although the amount of metals produced during this primordial phase is highly uncertain (Bromm and Larson 2004), our analysis limits it significantly and eases the requirement to satisfy the constraints set by the observations of the most metal-poor Population II stars.

\section{Conclusions}
\label{conc}

By making use of the available information on cosmic sources generating diffuse photons in the universe between the far-UV and the sub-millimeter over a wide range of cosmic epochs, we have estimated with the best possible time and spectral resolution the background photon density and its redshift evolution. This has exploited relevant data from ground-based observatories in the optical, near-IR, and sub-millimeter, as well as multi-wavelength information coming from space telescopes, namely HST, ISO and Spitzer. Additional constraints have been inferred from direct measurements or upper limits on the extragalactic backgrounds by dedicated missions (COBE).

Our estimated redshift-dependent background spectral shape has been used by us to evaluate the photon-photon opacity for sources of high-energy emission at any redshift. The same data can also be adopted to compute the optical depths for any kind of process in the intergalactic space involving interactions with background photons, like those of the (proton) cosmic rays generated by Active Nuclei and that are potentially responsible for high-energy neutrinos (Stanev et al. 2006).

Suitable tabulated forms are reported in the paper for the photon background (Tables \ref{tngamma} and \ref{tngamma1}), as well as for the photon-photon opacity corrections (Table \ref{tautab}).

A detailed account of evolutionary effects as a function of redshift in the infrared background photon densities is not needed to calculate the $\gamma-\gamma$ opacity in distant BLAZARs (from Fig. \ref{tauE} and eq. \ref{energy}, the background radiation at $\sim10\ \mu$m can be probed only up to z$\simeq$0.03 and that at 1.2 $\mu$m only up to z$\simeq$0.1). Photons emitted by high-z BLAZARs rather interact with optical-UV background light. 
However, evolutionary effects in the IR background light are critical for other reasons, such as the production of high-energy neutrinos by interactions of cosmic ray protons with IR photons, or the photo-excitation of molecular or atomic species in high redshift media.

We have applied our photon-photon opacity estimates to the analysis of spectral data at TeV energies on a few BLAZARs of particular interest.  The main results of these applications are summarized here.

   \begin{enumerate}
   
      \item 
In all  our analyzed cases, the opacity-corrected TeV spectra below $E_\gamma = 10 \ TeV$ are entirely consistent with standard photon-generation processes and all show photon indices steeper than $\Gamma_{intrinsic}=1.7$.
   
      \item 
The two local BLAZARs with spectral information above $E_\gamma = 10 \ TeV$, MKN 501 and MKN 421, have spectral shapes with $\Gamma_{intrinsic}\simeq 2$ and $\simeq 3$ independent of energy. There might be marginal indication that the opacity correction above 10 TeV might have been somewhat overestimated, which could have happened if our adopted photon background intensity at $\epsilon< 0.2\ eV$ ($\lambda > 10\ \mu$m) was a bit too large. However this conflicts with the fact that our knowledge of the cosmic sources emitting at 15 and 24 $\mu$m is quite precisely set by the deep ISO and Spitzer surveys, as summarized in Sect. \ref{model}, implying that our opacity corrections here are a reliable minimum. It will be interesting to check whether a conflict might arise with future observations of nearby sources at the highest photon energies with Cherenkov observatories.
   
      \item 
Contrary to previous claims, but in agreement with previous reports by Aharonian et al. (2006), we find no evidence for any truly diffuse background components in addition to those from resolved sources. 
We have tested in particular the effects of a background source originating at very high redshifts and completely unresolved by current instrumentation, such as Lyman photons from a primeval population of Population III stars emitting around $z\sim 10$. This background would be expected to have a peak redshifted emission between 1 and 2 $\mu$m locally. We could not identify any opacity features in our studied BLAZAR spectra consistent with such an emission and were able to set a stringent limit to such a diffuse photon intensity of $\sim 6 $  nW/m$^2$/sr between 1 and 4 $\mu$m, some 20\% of the IRTS detected signal.  This translates into a conservative limit of $f_\ast < 0.5\%$ of baryons to undergo primeval star-formation with maximal photon efficiency. However such emission is likely lower or even undetectable.

      \item 
 Detailed predictions for the extragalactic background radiation, and the photon-photon optical depths based on our modelling will be publicly distributed and kept updated on the WEB site
http://www.astro.unipd.it/background

   \end{enumerate}

%

   \begin{table*}
      \caption[]{Photon proper number density.}
         \label{tngamma}
\begin{tabular}{l l l l l l l l l l l l  }     
            \hline
            \noalign{\smallskip}
 \multicolumn{2}{c}{z=0}    & \multicolumn{2}{c} {z=0.2}   &
 \multicolumn{2}{c}{z=0.4}  & \multicolumn{2}{c} {z=0.6}   &   
 \multicolumn{2}{c}{z=0.8}  & \multicolumn{2}{c} {z=1.0}   \\
            \hline
$\log\epsilon$$^{\mathrm{a}}$&$\log(\epsilon\frac{dn_\gamma}{d\epsilon})$$^{\mathrm{b}}$   & $\log\epsilon$&$\log(\epsilon\frac{dn_\gamma}{d\epsilon})$   & $\log\epsilon$&$\log(\epsilon\frac{dn_\gamma}{d\epsilon})$   & $\log\epsilon$&$\log(\epsilon\frac{dn_\gamma}{d\epsilon})$   & $\log\epsilon$&$\log(\epsilon\frac{dn_\gamma}{d\epsilon})$   & $\log\epsilon$&$\log(\epsilon\frac{dn_\gamma}{d\epsilon})$   \\
            \noalign{\smallskip}
            \hline
            \noalign{\smallskip}
   -2.835  &  -0.9367 &  -2.756  &  -0.7423 & -2.689 &   -0.5311	&  -2.631 &   -0.3672 &  -2.58  &  -0.2245&  -2.534  &  -0.1062 \\
   -2.604  &  -0.3533 &  -2.525  &  -0.1642 & -2.458 &   0.04493	&    -2.4 &    0.2084 & -2.349  &   0.3406&  -2.303  &   0.4451 \\
    -2.45  &  -0.1126 &   -2.37  &  0.07846 & -2.303 &    0.2842	&  -2.245 &    0.4412 & -2.194  &   0.5546&  -2.148  &   0.6357 \\
   -2.303  & -0.03035 &  -2.224  &   0.1723 & -2.157 &    0.3701	&  -2.099 &    0.5147 & -2.048  &   0.6046&  -2.002  &   0.6636 \\
   -2.136  &  -0.0691 &  -2.057  &   0.1303 &  -1.99 &    0.3197	&  -1.932 &     0.449 & -1.881  &   0.5111&  -1.835  &   0.5391 \\
   -2.052  &  -0.1755 &  -1.972  &  0.01995 & -1.906 &    0.1956	&  -1.847 &    0.3064 & -1.796  &   0.3428&  -1.751  &   0.3508 \\
   -1.947  &  -0.3735 &  -1.868  &  -0.1889 & -1.801 &   -0.0317	&  -1.743 &   0.05994 & -1.692  &  0.08171&  -1.646  &  0.07664 \\
    -1.86  &  -0.5733 &   -1.78  &  -0.4037 & -1.714 &   -0.2571	&  -1.656 &   -0.1738 & -1.604  &  -0.1568&  -1.559  &  -0.1605 \\
    -1.78  &  -0.7724 &  -1.701  &  -0.6107 & -1.634 &   -0.4717	&  -1.576 &   -0.3885 & -1.525  &  -0.3697&   -1.48  &  -0.3711 \\
   -1.751  &  -0.8526 &  -1.671  &  -0.6934 & -1.604 &   -0.5541	&  -1.546 &   -0.4703 & -1.495  &  -0.4503&   -1.45  &  -0.4509 \\
   -1.684  &  -0.9834 &  -1.604  &  -0.8292 & -1.537 &   -0.6897	&   -1.48 &   -0.6048 & -1.428  &   -0.583&  -1.383  &   -0.583 \\
   -1.508  &   -1.418 &  -1.428  &   -1.265 & -1.361 &    -1.115	&  -1.303 &    -1.011 & -1.252  &  -0.9646&  -1.206  &   -0.942 \\
   -1.383  &   -1.634 &  -1.303  &   -1.469 & -1.236 &    -1.306	&  -1.178 &    -1.198 & -1.127  &   -1.146&  -1.082  &   -1.116 \\
   -1.303  &   -1.793 &  -1.224  &   -1.593 & -1.157 &    -1.430	&  -1.099 &    -1.317 & -1.048  &   -1.258&  -1.002  &   -1.241 \\
   -1.206  &   -1.883 &  -1.127  &   -1.688 &  -1.06 &    -1.517	&  -1.002 &    -1.410 &-0.9512  &   -1.348& -0.9055  &   -1.295 \\
   -1.082  &   -2.045 &  -1.002  &   -1.858 &-0.9355 &    -1.690	& -0.8775 &    -1.567 &-0.8262  &   -1.541& -0.7804  &   -1.576 \\
  -0.9846  &   -2.269 & -0.9055  &   -2.063 &-0.8386 &    -1.915	& -0.7804 &    -1.834 &-0.7293  &   -1.872& -0.6836  &   -1.935 \\
  -0.8598  &   -2.429 & -0.7804  &   -2.226 &-0.7135 &    -2.077	& -0.6556 &    -1.991 &-0.6045  &   -2.004& -0.5586  &   -2.038 \\
  -0.8085  &   -2.453 & -0.7293  &   -2.251 &-0.6623 &    -2.103	& -0.6045 &    -2.013 &-0.5533  &   -2.010& -0.5075  &   -2.026 \\
  -0.7314  &   -2.423 & -0.6523  &   -2.224 &-0.5854 &    -2.075	& -0.5274 &    -1.980 &-0.4763  &   -1.948& -0.4305  &   -1.939 \\
  -0.6688  &   -2.437 & -0.5897  &   -2.234 &-0.5227 &    -2.083	& -0.4647 &    -1.984 &-0.4136  &   -1.943& -0.3678  &   -1.929 \\
  -0.5597  &   -2.408 & -0.4795  &   -2.204 &-0.4136 &    -2.056	& -0.3556 &    -1.953 &-0.3044  &   -1.893& -0.2587  &   -1.866 \\
  -0.4713  &   -2.362 & -0.3921  &   -2.169 &-0.3251 &    -2.022	& -0.2672 &    -1.924 &-0.2160  &   -1.880& -0.1702  &   -1.872 \\
  -0.3792  &   -2.336 & -0.3000  &   -2.143 &-0.2331 &    -2.012	& -0.1751 &    -1.937 &-0.1240  &   -1.914&-0.07825  &   -1.927 \\
  -0.1359  &   -2.390 & -0.0567  &   -2.250 & 0.0100 &    -2.262	& 0.06819 &    -2.121 & 0.1193  &   -2.115&  0.1652  &   -2.138 \\
 -0.01936  &   -2.492 & 0.05195  &   -2.355 & 0.1268 &    -2.295	&  0.1847 &    -2.262 & 0.2358  &   -2.263&  0.2817  &   -2.283 \\
  0.09447  &   -2.638 &  0.1738  &   -2.535 & 0.2405 &    -2.468	&  0.2986 &    -2.441 & 0.3499  &   -2.451&  0.3955  &   -2.503 \\
   0.1915  &   -2.823 &  0.2707  &   -2.718 & 0.3377 &    -2.672	&  0.3955 &    -2.675 & 0.4467  &   -2.724&  0.4925  &   -2.799 \\
   0.3541  &   -3.141 &  0.4334  &   -3.105 & 0.5004 &    -3.093 	&  0.5583 &    -3.050 & 0.6095  &   -3.011&  0.6552  &   -2.963 \\
   0.4925  &    -3.54 &  0.5717  &   -3.440 & 0.6386 &    -3.352	&  0.6966 &    -3.236 & 0.7477  &   -3.148&  0.7935  &   -3.079 \\
   0.7935  &   -4.166 &  0.8727  &   -4.038 & 0.9396 &    -3.971	&  0.9976 &    -3.939 &  1.049  &   -3.921&   1.094  &   -3.879 \\
            \noalign{\smallskip}
            \hline
         \end{tabular}
\begin{list}{}{}
\item[$^{\mathrm{a}}$] Photon energies $\epsilon$ are in eV.
\item[$^{\mathrm{b}}$] Photon densities $\epsilon\frac{dn_\gamma}{d\epsilon}$ are in $[cm^{-3}]$.
\end{list}
   \end{table*}
%

   \begin{table*}
      \caption[]{Photon proper number density.}
         \label{tngamma1}
\begin{tabular}{l l l l l l l l l l }     
            \hline
            \noalign{\smallskip}
 \multicolumn{2}{c}{z=1.2}  & \multicolumn{2}{c} {z=1.4}   &
 \multicolumn{2}{c}{z=1.6}  & \multicolumn{2}{c} {z=1.8}   &
 \multicolumn{2}{c}{z=2.0}  \\
            \hline
$\log\epsilon$$^{\mathrm{a}}$&$\log(\epsilon\frac{dn_\gamma}{d\epsilon})$$^{\mathrm{b}}$   & $\log\epsilon$&$\log(\epsilon\frac{dn_\gamma}{d\epsilon})$   & $\log\epsilon$&$\log(\epsilon\frac{dn_\gamma}{d\epsilon})$   & $\log\epsilon$&$\log(\epsilon\frac{dn_\gamma}{d\epsilon})$   &
$\log\epsilon$&$\log(\epsilon\frac{dn_\gamma}{d\epsilon})$   \\
            \noalign{\smallskip}
            \hline
            \noalign{\smallskip}
   -2.492 & -0.009528	&   -2.455 &   0.06558	&   -2.42 &    0.1156	&  -2.388 &    0.1517	&   -2.358 &    0.1703 \\
   -2.262 &    0.5224	&   -2.224 &    0.5691	&  -2.189 &    0.5894	&  -2.157 &    0.5977	&   -2.127 &    0.5902 \\
   -2.107 &    0.6874	&   -2.069 &     0.703	&  -2.035 &     0.699	&  -2.002 &    0.6874	&   -1.972 &    0.6629 \\
   -1.961 &    0.6968	&   -1.923 &    0.6931	&  -1.888 &    0.6682	&  -1.856 &    0.6348	&   -1.826 &    0.5884 \\
   -1.793 &    0.5403	&   -1.756 &    0.5013	&  -1.721 &    0.4496	&  -1.689 &    0.3883	&   -1.659 &    0.3084 \\
   -1.709 &    0.3402	&   -1.671 &    0.2794	&  -1.637 &    0.2068	&  -1.604 &    0.1348	&   -1.574 &   0.05805 \\
   -1.604 &   0.05614	&   -1.567 & -0.003532	&  -1.532 &  -0.06778	&    -1.5 &   -0.1286	&    -1.47 &   -0.1946 \\
   -1.517 &    -0.177	&    -1.48 &   -0.2301	&  -1.445 &   -0.2892	&  -1.413 &   -0.3518	&   -1.383 &   -0.4279 \\
   -1.438 &   -0.3861	&     -1.4 &   -0.4394	&  -1.366 &   -0.4999	&  -1.333 &   -0.5591	&   -1.303 &   -0.6229 \\
   -1.408 &   -0.4646	&    -1.37 &   -0.5127	&  -1.336 &   -0.5672	&  -1.303 &   -0.6211	&   -1.273 &   -0.6834 \\
   -1.341 &   -0.5952	&   -1.303 &   -0.6364	&  -1.269 &   -0.6799	&  -1.236 &   -0.7261	&   -1.206 &   -0.7836 \\
   -1.165 &   -0.9404	&   -1.127 &   -0.9714	&  -1.093 &    -1.005	&   -1.06 &    -1.027	&    -1.03 &    -1.068 \\
    -1.04 &    -1.101	&   -1.002 &    -1.132	& -0.9678 &    -1.189	& -0.9355 &    -1.224	&  -0.9055 &    -1.227 \\
   -0.961 &    -1.235	&  -0.9234 &    -1.206	& -0.8884 &    -1.173	& -0.8564 &    -1.178	&  -0.8262 &     -1.239\\
  -0.8642 &    -1.253	&  -0.8262 &    -1.271	& -0.7916 &    -1.339	& -0.7595 &    -1.446	&  -0.7293 &    -1.563 \\
  -0.7392 &    -1.673	&  -0.7014 &    -1.808	& -0.6666 &    -1.907	& -0.6343 &    -1.981	&  -0.6045 &    -2.021 \\
  -0.6423 &    -1.973	&  -0.6045 &    -2.013	& -0.5698 &    -2.006	& -0.5375 &    -1.994	&  -0.5075 &    -1.999 \\
  -0.5173 &    -2.058	&  -0.4795 &    -2.084	& -0.4448 &    -2.078   & -0.4125 &    -2.067	&  -0.3826 &    -2.078 \\
  -0.4661 &    -2.036	&  -0.4283 &    -2.055	& -0.3936 &    -2.048	& -0.3614 &    -2.039	&  -0.3314 &    -2.053 \\
  -0.3891 &    -1.935	&  -0.3513 &    -1.944	& -0.3166 &    -2.048	& -0.2844 &    -1.930	&  -0.2545 &    -1.948 \\
  -0.3265 &    -1.907	&  -0.2887 &    -1.907	& -0.2539 &    -1.911	& -0.2217 &    -1.922	&  -0.1918 &    -1.951 \\
  -0.2173 &    -1.868	&  -0.1795 &    -1.891	& -0.1448 &    -1.921	& -0.1115 &    -1.956	& -0.08265 &    -2.005 \\
  -0.1289 &    -1.896	& -0.08155 &    -1.934  &-0.05637 &    -1.968	&-0.02462 &    -2.001	&  0.01536 &     -2.130 \\
 -0.04015 &    -1.952	&-0.002395 &    -1.985	& 0.03572 &    -2.015	& 0.06876 &    -2.053	&  0.09783 &    -2.103 \\
   0.2066 &     -2.16	&   0.2443 &    -2.175	&   0.272 &    -2.188	&  0.3042 &    -2.216	&   0.3342 &    -2.254 \\
    0.334 &    -2.321	&   0.3719 &    -2.349	&  0.3877 &    -2.345	&  0.4388 &    -2.434	&   0.4500 &    -2.466 \\
    0.433 &    -2.550	&   0.4708 &    -2.621	&  0.5095 &    -2.671	&  0.5378 &    -2.683	&   0.5717 &    -2.710 \\
   0.5339 &    -2.827	&   0.5717 &    -2.807	&  0.6064 &     -2.80	&  0.6386 &    -2.793	&   0.6686 &    -2.791 \\
   0.6966 &    -2.816	&   0.7344 &    -2.840	&  0.7692 &    -2.793	&  0.8013 &    -2.760	&   0.8313 &    -2.744 \\
   0.8349 &    -3.023	&   0.8727 &    -2.985	&  0.9075 &    -2.966	&  0.9396 &    -2.974	&   0.9696 &    -3.018 \\
    1.136 &    -3.777	&    1.174 &    -3.661	&   1.208 &    -3.573	&   1.241 &    -3.530	&    1.271 &    -3.501 \\
            \noalign{\smallskip}
            \hline
         \end{tabular}
\begin{list}{}{}
\item[$^{\mathrm{a}}$] Photon energies $\epsilon$ are in eV.
\item[$^{\mathrm{b}}$] Photon densities $\epsilon\frac{dn_\gamma}{d\epsilon}$ are in $[cm^{-3}]$.
\end{list}
   \end{table*}

   \begin{table*}
      \caption[]{Photon-photon optical depth as a function of energy and redshift$^{\mathrm{a}}$.}
         \label{tautab}
\begin{tabular}{l l l l l l l l l l  }     
            \hline
            \hline
            \noalign{\smallskip}
	Energy     & $\tau(z,E_\gamma)$ & $\tau(z,E_\gamma)$ & $\tau(z,E_\gamma)$ & $\tau(z,E_\gamma)$ & $\tau(z,E_\gamma)$ & $\tau(z,E_\gamma)$ & $\tau(z,E_\gamma)$ & $\tau(z,E_\gamma)$ & $\tau(z,E_\gamma)$  \\
            \noalign{\smallskip}
    [TeV]      &   z=0.01	 	 &	z=0.03	            &  z=0.1 		&	 z=0.3		    &   z=0.5       &   z=1.0	 &	   z=1.5    &  z=2.0	  &   z=3.0          \\
            \noalign{\smallskip}
            \hline
            \noalign{\smallskip}
 0.0200     & 0.0000        &   0.0000      	    & 0.0000     	& 0.000000          & 0.0000021  	&   0.004933 &	  0.0399	& 0.1157 	  &  0.2596 			  \\
 0.0240     & 0.0000        &   0.0000    	        & 0.0000     	& 0.000000 	    	& 0.000188  	&   0.01284  &	  0.0718	& 0.1783 	  &  0.3635 			  \\
 0.0289     & 0.0000        &   0.0000    	        & 0.0000     	& 0.000000   		& 0.001304   	&   0.0279 	 &	  0.1188	& 0.2598 	  &  0.4919 			  \\
 0.0347     & 0.0000        &   0.0000    	        & 0.0000     	& 0.000488  		& 0.004558  	&   0.0533 	 &	  0.1833	& 0.3635 	  &  0.6517 			  \\
 0.0417     & 0.0000         &   0.0000    		    & 5.254E-05 	& 0.002276  		& 0.01157   	&   0.0921 	 &	  0.2689	& 0.4967 	  &  0.8548 			  \\
 0.0502     & 0.0000         &   9.445E-05 		    & 5.408E-04 	& 0.006575   		& 0.02436   	&   0.1480 	 &	  0.3836	& 0.6745 	  &   1.118 			  \\
 0.0603     & 1.0976E-04     &   4.241E-04 		    & 1.915E-03 	& 0.014592  		& 0.04512   	&   0.2275 	 &	  0.5434	& 0.9179 	  &   1.465 			  \\
 0.0726     & 3.0882E-04     &   1.103E-03 		    & 4.548E-03 	& 0.02771   		& 0.07684    	&   0.3430 	 &	  0.7707	&  1.251 	  &   1.917 			  \\
 0.0873     & 6.5619E-04     &   2.258E-03 		    & 8.903E-03 	& 0.04808   		& 0.1248    	&   0.5137 	 &	   1.092	&  1.703 	  &   2.503 			  \\
 0.104      & 1.2130E-03     &   4.097E-03 		    & 1.582E-02 	& 0.07958    		& 0.1984    	&   0.7640 	 &	   1.537	&  2.302 	  &   3.249 			  \\
 0.126      & 2.1063E-03     &   7.039E-03 		    & 2.685E-02 	& 0.1284    		& 0.3109    	&    1.120 	 &	   2.133	&  3.073 	  &   4.181 			  \\
 0.151      & 3.5291E-03     &   1.167E-02 		    & 4.406E-02 	& 0.2031    		& 0.4780    	&    1.607 	 &	   2.905	&  4.042 	  &   5.318 			  \\
 0.182      & 5.7051E-03     &   1.872E-02 		    & 7.010E-02 	& 0.3134    		& 0.7163    	&    2.247 	 &	   3.875	&  5.225 	  &   6.673 			  \\
 0.219      & 8.9183E-03     &   2.907E-02 		    & 0.1082     	& 0.4696    		&  1.040    	&    3.056 	 &	   5.055	&  6.627 	  &   8.241 			  \\
 0.263      & 1.3517E-02     &   4.378E-02 		    & 0.1618     	& 0.6809    		&  1.461    	&    4.042 	 &	   6.438	&  8.226 	  &   9.997 			  \\
 0.316      & 1.9793E-02     &   6.367E-02 		    & 0.2338     	& 0.9517    		&  1.981    	&    5.192 	 &	   7.989	&  9.977 	  &   11.89 			  \\
 0.381      & 2.7938E-02     &   8.935E-02 		    & 0.3256     	&  1.281    		&  2.594    	&    6.474 	 &	   9.650	&  11.81 	  &   13.89 			  \\
 0.458      & 3.7957E-02     &   0.1205     		& 0.4356     	&  1.661    		&  3.284    	&    7.836 	 &	   11.34	&  13.67 	  &   15.93 			  \\
 0.550      & 4.9558E-02     &   0.1563     		& 0.5607     	&  2.082    		&  4.023    	&    9.214 	 &	   13.01	&  15.51 	  &   18.08 			  \\
 0.662      & 6.2291E-02     &   0.1953     		& 0.6961     	&  2.524    		&  4.779    	&    10.55 	 &	   14.63	&  17.39 	  &   20.45 			  \\
 0.796      & 7.5753E-02     &   0.2364     		& 0.8373     	&  2.967    		&  5.517    	&    11.82 	 &	   16.25	&  19.49 	  &   23.27 			  \\
 0.957      & 8.9194E-02     &   0.2768     		& 0.9750     	&  3.389    		&  6.210    	&    13.03 	 &	   18.04	&  22.02 	  &   26.81 			  \\
  1.15      & 0.1019    	 &   0.3152     		&  1.105     	&  3.779    		&  6.846    	&    14.29 	 &	   20.21	&  25.22 	  &   31.33 			  \\
  1.38      & 0.1136    	 &   0.3501     		&  1.223     	&  4.129    		&  7.432    	&    15.73 	 &	   22.98	&  29.37 	  &   37.23 			  \\
  1.66      & 0.1240    	 &   0.3810     		&  1.327     	&  4.444    		&  8.010    	&    17.54 	 &	   26.58	&  34.78 	  &   45.09 			  \\
  2.000     & 0.1329    	 &   0.4076     		&  1.419     	&  4.747    		&  8.652    	&    19.87 	 &	   31.31	&  41.95 	  &   55.80 			  \\
  2.40      & 0.1409    	 &   0.4318     		&  1.504     	&  5.079    		&  9.452    	&    22.96 	 &	   37.67	&  51.72 	  &   70.71 			  \\
  2.89      & 0.1486    	 &   0.4560     		&  1.596     	&  5.498    		&  10.52    	&    27.08 	 &	   46.30	&  65.17 	  &   92.14 			  \\
  3.47      & 0.1579    	 &   0.4863     		&  1.714     	&  6.075    		&  11.96    	&    32.66 	 &	   58.24	&  84.48 	  &   124.0 			  \\
  4.17      & 0.1710    	 &   0.5284     		&  1.879     	&  6.875    		&  13.92    	&    40.39 	 &	   75.45	&  113.2 	  &   172.1 			  \\
  5.02      & 0.1896    	 &   0.5887     		&  2.113     	&  7.952    		&  16.57    	&    51.39 	 &	   101.3	&  157.7 	  &   245.9 			  \\
  6.03      & 0.2162    	 &   0.6732     		&  2.431     	&  9.421    		&  20.24    	&    67.70 	 &	   141.4	&  227.3 	  &   357.0 			  \\
  7.26      & 0.2512    	 &   0.7847     		&  2.858     	&  11.45    		&  25.57    	&    92.73 	 &	   204.9	&  335.3 	  &   519.3 			  \\
  8.73      & 0.3017    	 &   0.9447     		&  3.464     	&  14.41    		&  33.52    	&    132.2 	 &	   304.3	&  496.4 	  &   747.0 			  \\
  10.5      & 0.3732    	 &    1.171     		&  4.334     	&  18.87    		&  45.81    	&    195.0 	 &	   454.1	&  724.1 	  &   1048. 			  \\
  12.6      & 0.4795    	 &    1.513     		&  5.663     	&  25.76    		&  65.21    	&    292.5 	 &	   666.6	&  1027. 	  &   1426. 			  \\
  15.2      & 0.6455    	 &    2.048     		&  7.723     	&  36.60    		&  95.98    	&    435.4 	 &	   948.5	&  1407. 	  &   1873. 			  \\
  18.2      & 0.8984    	 &    2.871     		&  10.93     	&  53.79    		&  143.7    	&    630.5 	 &	   1299.	&  1852. 	  &   2372. 			  \\
  21.9      &  1.297    	 &    4.162     		&  15.99     	&  80.41    		&  214.3    	&    878.2 	 &	   1705.	&  2339. 	  &   2897. 			  \\
  26.4      &  1.917    	 &    6.177     		&  23.86     	&  119.9    		&  311.5    	&    1172. 	 &	   2145.	&  2843. 	  &   3412. 			  \\
  31.7      &  2.856    	 &    9.181     		&  35.47     	&  174.8    		&  435.3    	&    1494. 	 &	   2587.	&  3323. 	  &   3887. 			  \\
  38.1      &  4.211    	 &    13.47     		&  51.78     	&  245.0    		&  580.9    	&    1824. 	 &	   3004.	&  3752. 	  &   4303. 			  \\
  45.8      &  6.038    	 &    19.14     		&  72.76     	&  327.4    		&  739.9    	&    2134. 	 &	   3362.	&  4102. 	  &   4618. 			  \\
  55.1      &  8.285    	 &    26.00     		&  97.51     	&  416.9    		&  899.0    	&    2403. 	 &	   3638.	&  4352. 	  &   4835. 			  \\
  66.2      &  10.82    	 &    33.59     		&  124.3     	&  506.3    		&  1045.    	&    2616. 	 &	   3825.	&  4499. 	  &   4940. 			  \\
  79.6      &  13.48    	 &    41.42     		&  151.2     	&  587.7    		&  1169.    	&    2756. 	 &	   3915.	&  4540. 	  &   4945. 			  \\
  95.7      &  16.04    	 &    48.81     		&  175.8     	&  655.4    		&  1263.    	&    2823. 	 &	   3915.	&  4540. 	  &   4945. 			  \\
  115.      &  18.24    	 &    54.98     		&  195.8     	&  705.7    		&  1320.    	&    2823. 	 &	   3915.	&  4540. 	  &   4945. 			  \\
  138.      &  20.01    	 &    59.82     		&  210.8     	&  735.5    		&  1340.    	&    2823. 	 &	   3915.	&  4540. 	  &   4945. 			  \\
  166.      &  21.20    	 &    63.03     		&  219.8     	&  744.0    		&  1340.    	&    2823. 	 &	   3915.	&  4540. 	  &   4945. 			  \\
            \hline
            \noalign{\smallskip}
            \hline
         \end{tabular}
\begin{list}{}{}
\item[$^{\mathrm{a}}$] The absorption multiplicative factor is $e^{-\tau(z,E_\gamma)}$.
\end{list}
   \end{table*}
%


\begin{acknowledgements}
      Part of this work was supported by the Italian Space Agency
under contract  ASI/INAF I/005/07/0 Herschel "Fase E". We warmly thank T. Stanev, A. Saggion, D. Mazin, M. Persic, and H. Krawczynski for useful comments and suggestions. We are also grateful to the anonymous referee for a careful reading of the original manuscript and useful criticisms.
\end{acknowledgements}

\end{document}